\documentclass[reprint,twocolumn, superscriptaddress, showkeys, nolongbibliography]{revtex4-1}
\usepackage{amsmath} 
\usepackage{graphicx}
\usepackage{dcolumn}
\usepackage{bm}
\usepackage{amsthm}
\usepackage{amsfonts}
\usepackage{xcolor}

\newcommand{\SIfigone}{1}
\newcommand{\SIfigtwo}{2}
\newcommand{\SIfigthree}{3}
\newcommand{\SIfigfour}{4}
\newcommand{\SIfigfive}{5}
\newcommand{\SIfigsix}{6}
\newcommand{\SIfigseven}{7}
\newcommand{\SIfigeight}{8}
\newcommand{\EQ}{\begin{equation}}
\newcommand{\EE}{\end{equation}}
\newcommand{\EQA}{\begin{eqnarray}}
\newcommand{\EEA}{\end{eqnarray}}
\begin{document}

\preprint{AIP/123-QED}

\title{Rational T~cell repertoire allocation to control the gut microbiota} 

\author{Obinna A. Ukogu}
\email{oukogu@stanford.edu.}
\affiliation{Department of Applied Mathematics, University of Washington, Seattle, WA, USA}
\affiliation{Current address: Department of Applied Physics, Stanford University, Stanford, CA, USA}
\author{Angelina M. Bilate}%
\affiliation{Laboratory of Mucosal Immunology, The Rockefeller University, New York, NY, USA}
\author{Thierry Mora}%
\affiliation{Laboratoire de Physique de l'\'Ecole Normale Sup\'erieure, CNRS, PSL University, Sorbonne University,
Universit\'e Paris-Cit\'e, Paris, France}
\author{Aleksandra M. Walczak}%
\affiliation{Laboratoire de Physique de l'\'Ecole Normale Sup\'erieure, CNRS, PSL University, Sorbonne University,
Universit\'e Paris-Cit\'e, Paris, France}
\author{Armita Nourmohammad}%
\email{armita.nourmohammad@yale.edu}
\affiliation{Department of Applied Mathematics, University of Washington, Seattle, WA, USA}
\affiliation{Department of Physics, University of Washington, Seattle, WA, USA}
\affiliation{Allen School of Computer Science and Engineering, University of Washington, Seattle, WA, USA}
\affiliation{Herbold Computational Biology Program, Fred Hutchinson Cancer Center, Seattle, WA, USA}
\affiliation{Center for Systems and Engineering Immunology, and the Departments of Immunobiology, Biomedical Engineering, and Physics, Yale University, New Haven, CT, USA}

\date{\today}

\begin{abstract}
The gut microbiota shapes host metabolism, development and immunity, yet how the host regulates its composition in the presence of complex ecological interactions remains poorly understood. Recent experiments that measured T~cell reactivity to individual strains in germ-free mice inoculated with a defined microbial community found that T~cell recognition across microbial strains is sparse but broad, with a few polyspecific clones recognizing antigens shared across taxonomically diverse strains. Reanalyzing these experiments, we show that these T~cell responses are associated with abundance changes of the strains they target. To interpret these findings, we develop a model linking the T~cell repertoire to microbial abundances, in which a repertoire is allocated to minimize microbe-driven harm at the cost of maintaining an immune response. The model rationalizes the observed sparsity of gut T~cell reactivity and shows that this allocation can emerge from competition between T~cells for microbe-derived antigens. Finally, we find that while immune recognition can regulate microbial abundances, ecological interactions between strains may interfere with this mechanism. Together, these results provide a quantitative framework for immune control of gut microbial composition and identify principles governing the organization of T~cell repertoires in the gut.
\end{abstract}

\maketitle

\section{Introduction}

The gut microbiome is a complex ecological community that influences its host's metabolism, development, immunity and other critical physiological processes~\cite{Hooper2012-el, Ansaldo2021-mh, Arnoldini2025-my, Palm2014-bz, Yamada2025-nd, Wang2023-tc}. Given the easy accessibility of the microbiome through the intestinal tract, it is a promising delivery route for clinical therapies, such as antibiotics, probiotics, and fecal matter transplants. However, the variability of microbiota composition among healthy individuals makes it difficult to identify stable drivers of health and disease~\cite{Wang2023-tc, Schmitt2026-we, Skwara2023-mb}. Recently, it was reported that despite compositional variation, the production of many metabolites is stable across healthy individuals on similar diets~\cite{Arnoldini2025-my}. Other studies found that host-secreted antibody-coating of strains can reliably identify colitogenic, harmful strains in the gut~\cite{Palm2014-bz, Vujkovic-Cvijin2022-wh}. These examples suggest certain properties of the microbiota---metabolism, and perhaps immunity---are stably regulated by healthy hosts. Still, given the complex ecological interactions in the gut, it is unclear how host health can be reliably influenced by perturbing microbiota composition. Understanding the logic behind host-control of microbiota composition could thus enhance the effectiveness of clinical therapies~\cite{Atarashi2013-re,Stein2018-dt, Tanoue2019-at,Littmann2021-om, Yeh2024-tz, Ryan2025-sz, Lentsch2025-aq, Huus2026-zw}.

\begin{figure*}[ht!]
\centering
\includegraphics[width=0.8\linewidth]{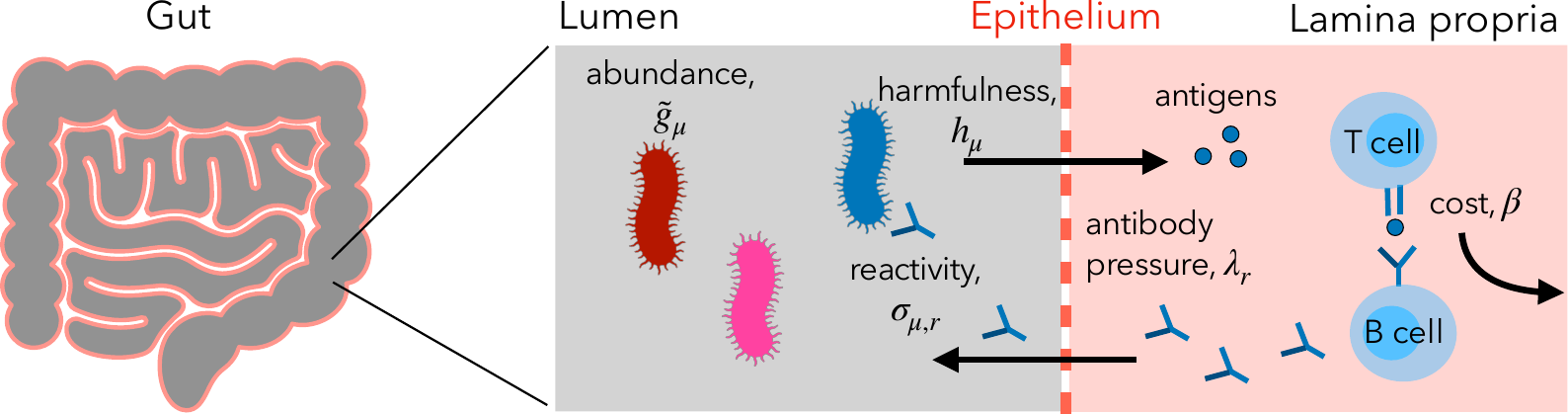}
\caption{{\bf A model of adaptive immune control of the gut microbiota.} Microbial abundance in the gut lumen ($\tilde{g}_\mu$) is influenced, among other factors, by microbiota ecology and host immunity. Microbes have distinct degrees of harmfulness ($h_\mu$) reflecting their propensity to secrete harmful metabolites, damage tissues or cross the epithelium. As strains inflict harm at rate ($h_\mu\tilde{g}_\mu$) and potentially trigger infection, they stimulate cognate T~cells that can bind to their antigens. T~cells direct B cells to secrete antibodies (with immune pressure $\lambda_r$) into the lumen, which bind specifically (with reactivity $\sigma_{\mu,r}$) to strains and decrease their abundance. The T-cell-dependent immune response may also trigger off-target harms to the host at a marginal rate $\beta$.}
\label{fig1}
\end{figure*}

Few quantitative models have linked the immune response to microbiota composition in the gut~\cite{Stein2018-dt}. In the digestive tract, environmental and dietary factors determine which strains are likely to be present, limiting the host's control of microbiota composition. Colonizing strains become entrenched by attaching to the gut epithelium and may harm their host by secreting harmful metabolites and damaging tissues (Fig. \ref{fig1}). Hosts, in turn, influence the abundance of harm-inducing microbes by modulating their internal environment through metabolism and immunity, not unlike the action of clinical pro- and anti-biotic treatments. The immune response recognizes specific microbial antigens to modulate microbe growth. It can inhibit growth through the secretion of bactericidal molecules and antibodies, or tolerate strains to maintain their abundance~\cite{Hooper2012-el, Palm2014-bz, Olm2025-sy}.The balance between inhibition and tolerance mechanisms directed towards a microbe is thought to influence its abundance in the gut, with certain strains reliably inducing inhibitory or tolerogenic immune responses~\cite{Atarashi2013-re, Klawon2025-hw, Su2016-zo, Hofer2012-qm, McNally2011-mh, Humrich2010-wc, Littmann2021-om, Schirmer2016-ch, Lisicka2025-ls, Vujkovic-Cvijin2022-wh, Moor2017-jv, Gu2024-gw, Olm2025-sy, Palm2014-bz, Probstel2020-pa}. Poorly managed allocation of immune coverage risks triggering runaway immune responses, leading to dysbiosis, allergies or other pathologies~\cite{Hooper2012-el, Kato2014-ph, Littmann2021-om, Tanoue2019-at, Geva-Zatorsky2017-ad, Derosa2022-ei, Yeh2024-tz, Wastyk2021-sq, Sanchez-Martinez2025-qd}.

It is unclear how immune mechanisms control microbiota composition given the complexity of the gut ecology. Here, we develop a quantitative model linking the T~cell repertoire to microbial composition in the gut. Our approach is motivated by, and leverages data from, recent experiments that measured immune interactions with strains in germ-free mice inoculated with a defined microbial community~\cite{Nagashima2023-fe, Cheng2022-mv, Wang2023-tc}. We find that the immune response partly determines microbial abundances, but that other factors, such as ecological interactions between strains, may be significant. Our model further rationalizes the experimentally observed pattern of sparse yet broad T~cell microbe interactions, showing it to be optimal for microbial control when maintaining an immune repertoire is costly. We demonstrate that this optimal repertoire allocation is biologically plausible and can emerge from competition between T~cells for microbe-derived antigens. Finally, we propose experiments to further refine the principles governing immune control of the gut microbiota composition.

\section{Model}
We introduce a model, in which a T-cell repertoire is allocated to minimize the harm from a microbiota. Each microbe (labeled~$\mu = 1,\dots,M$) engages with the immune system through its immune coverage:

\begin{align}
C_\mu(\vec\lambda) = \sum_{r =1}^R \sigma_{\mu,r}\lambda_r
\label{eq:coverage},
\end{align}
where $\lambda_r~(\text{for }r = 1,\dots, R)$ is the inhibitory immune pressure exerted by a T~cell carrying receptor $r$, and $ \sigma_{\mu,r}$ is the dimensionless reactivity between microbe $\mu~$ and the immune receptor $r$, defining a non-negative $M\times R$ reactivity matrix $\Sigma = [\sigma_{\mu,r}]$ (Fig. \ref{fig1}). We define the immune pressure of a T~cell clone as the relative abundance of its effector and regulatory T cell fates ($\lambda_r \sim |\text{Teff}|/|\text{Treg}|$). Immune pressure acts on microbes by influencing the concentration of the T-cell-dependent, high-affinity antigen-specific immunoglobulin A (IgA) secreted into the lumen by B~cells at the gut mucosal surface~\cite{Slack2012-nc, Pabst2012-rp, Palm2014-bz}. These antibodies bind to targeted strains to inhibit their growth. The coverage $C_\mu$ thus reflects the cumulative inhibitory pressure from all such antibodies generated by the repertoire targeting detectable epitopes of microbe $\mu$.

By exploiting niches in the gut, each microbe $\mu$ can grow to a carrying capacity $g_\mu^\text{max}$ that is modulated by its immune coverage $C_\mu(\vec\lambda)$. Since immune repertoire dynamics are slow (weeks) relative to microbial turnover (days)~\cite{Palm2014-bz, Cheng2022-mv}, we model only the steady-state microbe abundances $\tilde{g}_\mu$ (for $\mu = 1,\dots,M$). Neglecting potential ecological interactions between strains, we posit the simple relationship for the steady-state microbial abundances:

\begin{align}
\label{eq:microbe-steady-noecology}
\begin{split}
\tilde{g}_\mu(\vec\lambda) &= \frac{g_\mu^\text{max}}{1 + C_\mu(\vec\lambda)/K_C},
\end{split}
\end{align}
where $K_C$ is the half-maximum constant, that is, the magnitude of coverage at which microbe abundance is reduced to half its maximum. Each microbe $\mu$ inflicts systemic harm on the host at rate $h_\mu\tilde{g}_\mu$, where the constant $h_\mu >0$ quantifies its harmfulness, for example through its propensity to secrete harmful metabolites, damaging tissues or cross the epithelium \cite{Vujkovic-Cvijin2022-wh, Palm2014-bz, Saleh2011-tx}. Strains with low $h_\mu$ correspond to commensals while strains with high $h_\mu$ are pathobionts.

We assume that the repertoire response can minimize harm by controlling the steady-state population of microbes. Given the reactivity matrix $\Sigma = [\sigma_{\mu,r}]$ between microbial strains ($\mu$) and immune receptors ($r$), we seek a repertoire that minimizes the objective,

\begin{align}
\label{eq:cost_function}
\min_{\vec\lambda \geq 0}\left\{ \sum_{\mu =1}^M h_\mu \tilde{g}_\mu(\vec\lambda) + \beta\sum_{r = 1}^R \lambda_r \right\}.
\end{align}
The first term on the right represents the cumulative steady-state rate of harm inflicted by the microbiota. The second term captures the cost of maintaining the immune repertoire, associated with factors such as increased risk of immunopathology and autoimmunity in more effector-rich tissues, the difficulty of generating particular antigen-specific receptors, which may be rare in the na\"ive pool, and competition among receptor-bearing cells for limited survival signals and metabolic resources ~\cite{Hooper2012-el, Yeh2024-tz,Ganeshan2019-zq} (Fig. \ref{fig1}). The linear form of the second term, chosen for simplicity, corresponds to a sparsity promoting L1 regularizer~\cite{Boyd2004-qp}, with the constant $\beta$ specifying the marginal cost of immune pressure. Solutions of this optimization problem can be compared to features of the experimentally observed CD4+ T~cell repertoire in the gut.

\section{Results}

\subsection{Sparse, polyspecific T~cell coverage of the gut microbiota.}

\begin{figure*}[t!]
\centering
\includegraphics[width=0.8
\linewidth]{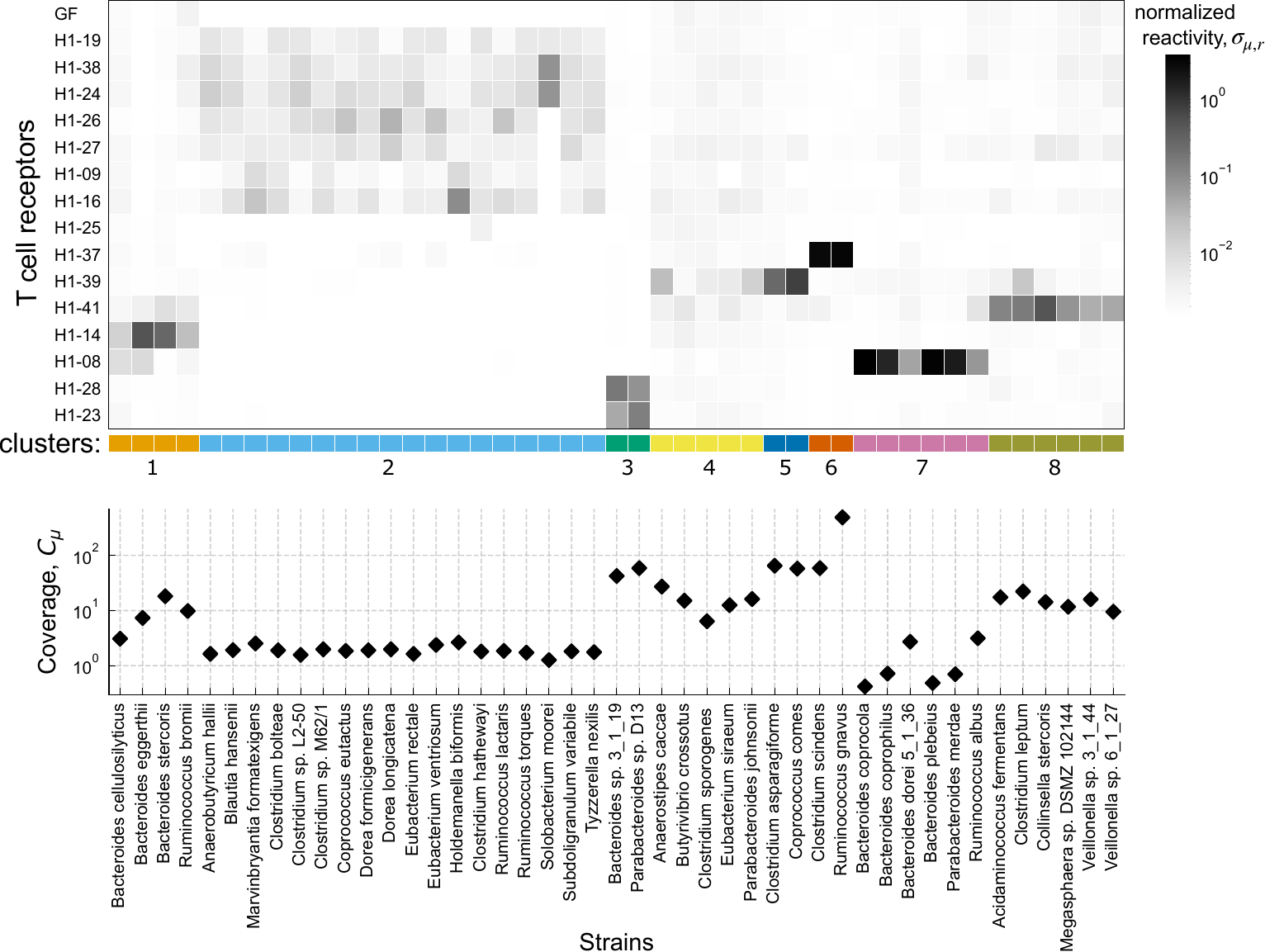}
\caption{{\bf T~cell repertoire reactivity and coverage in a complex microbial community.} Top: Heatmap showing strain reactivity (secreted IL-2 concentrations; L-1 normalized columns) of 45 microbial strains from a 97-member community (hCom1) to 15 distinct T~cell receptors; reproduced from ref.~\cite{Nagashima2023-fe}. The ``GF'' row provides a baseline for non-specific responses. K-means clustering based on these reactivities identifies 8 reactive groups (clusters 1-8), while 52 non-reactive strains (cluster 0) are omitted. Bottom: Estimated repertoire coverage ($C_\mu$) per strain calculated as a reactivity-weighted ratio of effector-to-regulatory cells (eq. \ref{eq:approx-immune-coverage}). Coverage varies significantly across strains, with potential pathobionts receiving higher coverage than commensals.
}
\label{fig2}
\end{figure*}

In recent experiments, germ-free mice were inoculated with a 97-member microbial community (``hCom1"). In one set of studies, they assessed metabolic output and stability of colonization to obtain compositional time series data~\cite{Cheng2022-mv, Wang2023-tc}. In another, they profiled T~cells isolated from the lamina propriae of the small and large intestines two weeks post-inoculation to obtain immune coverage data~\cite{Nagashima2023-fe}. Responding T~cell receptors found in the lamina propria were sequenced and profiled for their reactivity to individual strains in the community, generating an empirical reactivity matrix (Fig. \ref{fig2}, \textit{top}). Forty-five of these strains elicited a response, and the remaining 52 strains were nonreactive to all sampled clones and were excluded from the analysis. A majority of sampled clones were nonreactive to microbial strains, yet a subset recognized food- or microbe-derived antigens. Furthermore, taxonomically related microbial strains, which are expected to express similar antigens, were often targeted by distinct T~cell clones, while taxonomically diverse strains were often targeted by the same T~cell clone. Thus, this sample consisted of polyspecific T~cells, which sparsely recognized many taxonomically diverse sets of strains.

A given clone contained subsets of regulatory (Treg: tTreg and pTreg), effector (Teff: Th1, Th2, Th17, other effectors), memory, and additional T~cell phenotypes (SI Appendix, Fig. \SIfigone). Here, we focus on Treg and Teff subsets as we are interested in their opposing contributions to immune-modulatory responses (immune coverage) in the gut. To evaluate microbe-specific immune coverage, we analyzed the empirically measured reactivity matrix $\Sigma = [\sigma_{\mu,r}]$ between strains $\mu$ and receptors $r$ (Fig. \ref{fig2}, \textit{top}) for $R=15$ T-cell clones against $M=97$ microbial strains~\cite{Nagashima2023-fe}. Assuming each clone recognizes a specific microbe-expressed antigen through its unique TCR $r$, the balance of effector to regulatory cells within that clone reflects the net immune pressure it exerts on its target~\cite{Cong2009-pb, Atarashi2013-re, Yang2014-ip,Klawon2025-hw, Su2016-zo, Hofer2012-qm, McNally2011-mh, Humrich2010-wc, Littmann2021-om}. To avoid possible artifacts from pooling cells from different mice and noisy Teff-to-Treg ratio estimates, we combined cell counts across clones before taking the ratio, approximating the relative repertoire coverage of a strain $\mu$ (eq.~\ref{eq:coverage}) as

\begin{align}
\label{eq:approx-immune-coverage}
  C_\mu \approx \sum_{r=1}^{R}\sigma_{\mu,r}|\text{Teff}|\Bigg/\sum_{r=1}^{R}\sigma_{\mu,r}|\text{Treg}|,
\end{align}
where $|\cdot|$ represents the number of observed Teff/Treg cells. This expression retains the interpretation of coverage as a reactivity-weighted Teff-to-Treg ratio (eq.~\ref{eq:coverage}) while reducing sensitivity to sparsely sampled clones (SI Appendix, Fig.~\SIfigone). Using this measure, we found that coverage varied substantially across strains (Fig. \ref{fig2}, \textit{bottom}). For instance, \textit{Bacteroides coprocola}, a commensal, exhibited the lowest coverage, while \textit{Ruminococcus gnavus}, a potential pathobiont, exhibited the highest coverage~\cite{Liu2026-oo, Henke2019-ry}.

\subsection{Targeted repertoire coverage controls microbiota composition.}
 
\begin{figure*}[ht!]
\centering
\includegraphics[width=.8\linewidth]{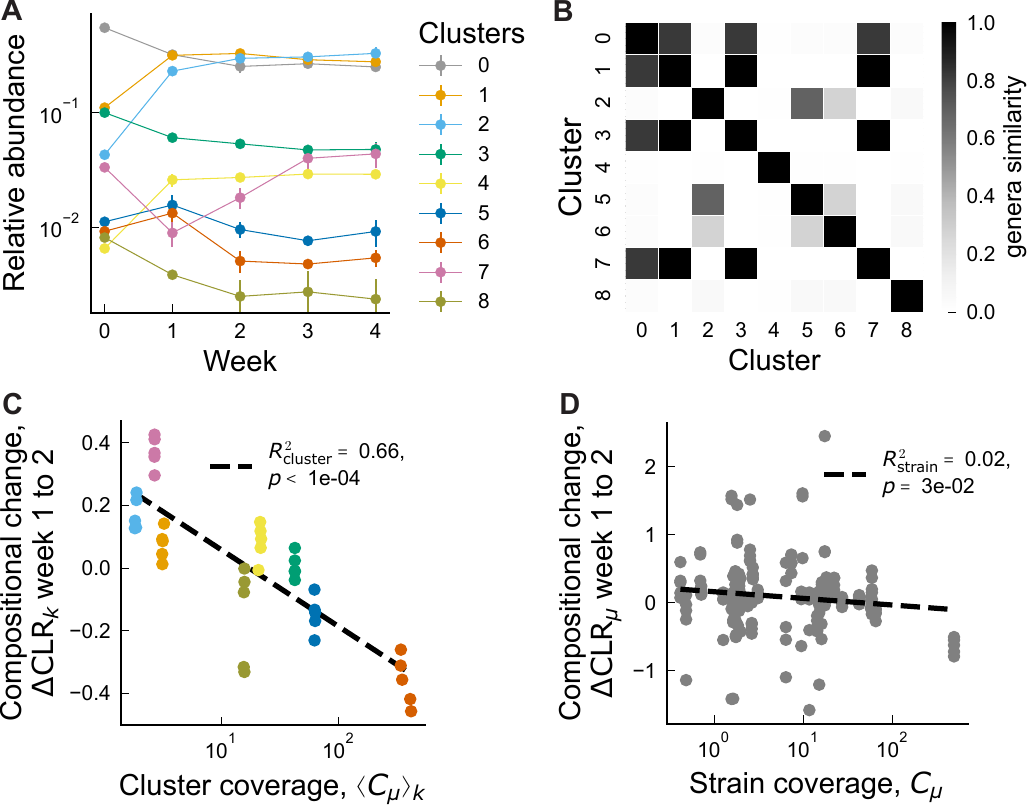}
\caption{{\bf Repertoire coverage predicts compositional shifts of reactive strain clusters.} {\bf(A)} Longitudinal tracking of cluster relative abundances in the entire microbiota over 4 weeks following inoculation. Error bars indicate 1 standard deviation across 5 biological replicates. {\bf(B)} Bray-Curtis similarity heatmap showing low taxonomic (genera) overlap between all 9 clusters. {\bf(C)} Scatter plot showing relationship between the average repertoire coverage of a given cluster $k$ ($\langle C_\mu\rangle_k$) and the change in its centered log-ratio ($\Delta\text{CLR}_k$; eq. \ref{eq:cluster_clr}) from week 1 to week 2 ($R^2_\text{cluster} = 0.66$). Colors are as in (A) and points of the same color are biological replicates. Dotted line represents fit to $\Delta\text{CLR}_k = m\log_{10}(\langle C_\mu \rangle_k) + b$ with $m=-0.25$, and $b=0.3$. Statistical significance was quantified through a hierarchical permutation test by shuffling cluster and biological replicate labels $n = 10000$ times while preserving the grouping of replicates. {\bf(D)} No relationship between individual strain composition changes ($\Delta\text{CLR}_\mu$) from week 1 to 2 and repertoire coverage ($C_\mu$; $R^2_\text{strain} = 0.02$). Points indicate strains and their biological replicates. Dotted line represents fit to $\Delta\text{CLR}_\mu = m\log_{10}(C_\mu) + b$. Statistical significance was quantified as in (C) at the strain level.
}
\label{fig3}
\end{figure*}

Recent research has shown that clustering strains based on their genetic similarity can predict the metabolic outputs of complex communities~\cite{Gowda2022-bh,Skwara2023-mb}. Inspired by our model formulation, we reasoned that grouping strains by their reactivity to T~cells ($\Sigma =[\sigma_{\mu,r}]$) could similarly reveal how collective immune coverage shapes community dynamics. We applied $K-$means clustering to the reactivity profiles of each strain, identifying nine microbial clusters in hCom1 (clusters 0-8; Fig. \ref{fig2}, \textit{middle}; SI Appendix). Strains in cluster 0 (52 total) were nonreactive to experimentally extracted T~cells, while those in clusters 1-8 (45 total) were reactive to at least one T~cell clone~\cite{Nagashima2023-fe}. Clusters varied in their relative abundance and exhibited distinct and reproducible compositional dynamics (Fig. \ref{fig3}A; SI Appendix, Fig.~\SIfigtwo), even if they contained multiple taxa (Fig. \ref{fig2}). Strains of the same taxon were often dispersed across different clusters (Fig. \ref{fig2}; SI Appendix, Fig.~\SIfigtwo), and clusters were dissimilar in taxon membership when measured by their overlap in strain genera (Fig. \ref{fig3}B). Such immune-based clustering is distinct from taxonomic relationships between strains and defines an aggregation of taxonomically diverse strains in the microbiota based on their potential immune responses.

In~vitro, the composition of hCom1 stabilizes within 48 hours in the absence of perturbations~\cite{Cheng2022-mv}, while in mice inoculated with different communities, the T~cell response takes weeks to develop~\cite{Palm2014-bz}. We use the quantities defined by our model to interpret changes in the microbial composition. Changes from week 0 to week 1 likely reflect microbial ecology and innate immune responses alone, and changes from week 1 to subsequent weeks are additionally shaped by the adaptive immune response. First, we define the repertoire coverage of the microbial cluster $k$, as the strain-abundance-weighted average immune coverage of the strains in each cluster:

\begin{align}
\langle C_\mu\rangle_k = \sum_{\mu \in\text{cluster}\, k} \frac{g_\mu}{{\sum\limits_{\mu \in\text{cluster}\, k} }g_\mu}C_\mu,\label{eq:cluster_coverage}
\end{align}
which is computed using the measured relative abundance $g_\mu/\sum_\alpha g_\alpha$ at week 1 (SI Appendix, Fig.~\SIfigtwo).
Next, we quantify cluster-level compositional shifts in the microbiota by computing the centered log-ratio (CLR) transform of relative abundances of the clusters at each time $t>0$. For each cluster $k=0,\dots,8$, we have

 \begin{align}
 \begin{split}
   \text{CLR}_k{(t)} = \log_{10}\left(G_k(t)\Bigg/\sqrt[9]{\prod_{j = 0}^{8} G_j(t)}~\right),\\ 
 \end{split}
 \label{eq:cluster_clr}
 \end{align}
where $ G_k(t) = \sum_{\mu\in \text{cluster} \,k}g_\mu(t)\Big/\sum_{\mu =1}^Mg_\mu(t)$ is the relative abundance of cluster $k$. This transformation standardizes the $M$ microbial strain trajectories, removing spurious correlations caused by the sum-to-one constraint of relative abundance data. CLR for individual strains can be computed analogously (SI Appendix) and span 8 decades from the most to least abundant strain. Changes in abundance were then computed between sampled time points ($t_1$ to $t_2$) following inoculation, $\Delta\text{CLR}_k(t_1,t_2)$. 

Plotting the compositional changes $\Delta \text{CLR}_k(1,2)$ of reactive clusters, $k=1,\dots,8$, from week 1 to week 2 shows a strong inverse relationship with log-cluster coverage $\log \langle C_\mu\rangle_k$ ($R^2_\text{cluster} = 0.66$; Fig. \ref{fig3}C), indicating that higher immune coverage more strongly constrains the abundance of a strain cluster. We excluded cluster 0 as its coverage is undefined. The relationship between compositional changes and immune coverage persisted when comparing week 1 to weeks 3 and 4 and was robust to dropping individual clusters or biological replicates (SI Appendix, Fig.~\SIfigthree). Additionally, there was not a significant correlation between abundance at week 1 and compositional change from week 1 to week 2 ($R^2 = 0.08$; SI Appendix, Fig.~\SIfigthree). Notably, clusters 1 and 2 which are the most abundant clusters at week 1 have the lowest coverages and the largest CLR increases (Fig. \ref{fig3}A). This is consistent with previous reports that strains in these clusters are likely commensals~\cite{Nagashima2023-fe}, and so may not require extensive immune coverage. The predominant ($>60\%$) strains in these clusters are \textit{Bacteroides cellulosilyticus} (commensal) and \textit{Clostridium hathewayi} (commensal/opportunistic), respectively~\cite{Robert2007-ig, Elsayed2004-lv}. Cluster 6 has the highest immune coverage and the largest decrease in CLR abundance (Fig. \ref{fig3}C). This cluster contains \textit{R. gnavus}, which has been associated with inflammatory bowel disease (IBD)~\cite{Henke2019-ry}. Interestingly, high-coverage clusters were suppressed but not eliminated (Fig. \ref{fig3}C), and the inverse relationship observed at the cluster level is absent at the individual strain level (Fig.~\ref{fig3}D), suggesting limits or diminishing returns to immune coverage which we investigate later in this work. Together, these results indicate that T~cell repertoire coverage distinguishes commensal from harmful microbial clusters and shapes microbiota composition accordingly~\cite{Atarashi2013-re, Schirmer2016-ch, Lisicka2025-ls, Vujkovic-Cvijin2022-wh, Moor2017-jv, Gu2024-gw, Olm2025-sy, Palm2014-bz, Probstel2020-pa}.

\subsection{Model of immune repertoire allocation predicts sparse, polyspecific gut T~cell repertoires}

Next, we asked if our model could explain three features of observed repertoires in the gut: the relative sparsity of T cell reactivity (52/97 nonreactive clones), the prevalence of polyspecific T cell clones (Fig. \ref{fig2}, \textit{top}) and the variability in coverage $C_\mu$ of microbial strains (Fig. \ref{fig2}, \textit{bottom}). To build intuition, we first consider the simplest case where each immune receptor recognizes one microbe and each microbe is recognized by one receptor. Following (eq.~\ref{eq:microbe-steady-noecology}), for a given microbe-receptor pair with reactivity $\sigma\geq 0$, repertoire coverage and microbe abundance are given by $C=\sigma \lambda$ and $\tilde{g}(\lambda) = g^\text{max}(1 +\sigma\lambda/K_C)^{-1}$. The optimal solution which minimizes the costs of microbial pathogenicity and immune response in the host (eq.~\ref{eq:cost_function}) satisfies,

\begin{align}
\label{eq:one-type-optimal-immune-coverage}
 C^* = \max\left[\Omega^{\frac{1}{2}} - 1,~ 0\right]\cdot K_C,\, \text{ with}\quad \Omega = \frac{\sigma hg^\text{max}}{K_C\beta}
\end{align}
where $\Omega$ is the protection-to-cost ratio, determined by the harmfulness of the microbe $h > 0$, the marginal cost of a receptor $\beta$, and the microbe carrying capacity $g^\text{max}$. Consistent with recent experiments~\cite{Ansaldo2019-fj, Yang2020-bp}, the model predicts that perturbing the carrying capacity (or abundance) of an immunogenic strain will shift its immune coverage nonlinearly. Similar square-root dependencies appeared in models of optimal immune memory allocation and optimal security screening~\cite{Press2009-fu, Mayer2015-ii}. In those contexts, the scaling was interpreted as flattening the distribution of allocated memory/surveillance. Here, it reflects the diminishing returns of marginal increases of repertoire coverage (eq. \ref{eq:microbe-steady-noecology}).

\begin{figure*}[ht!]
\centering
\includegraphics[width=0.9\linewidth]{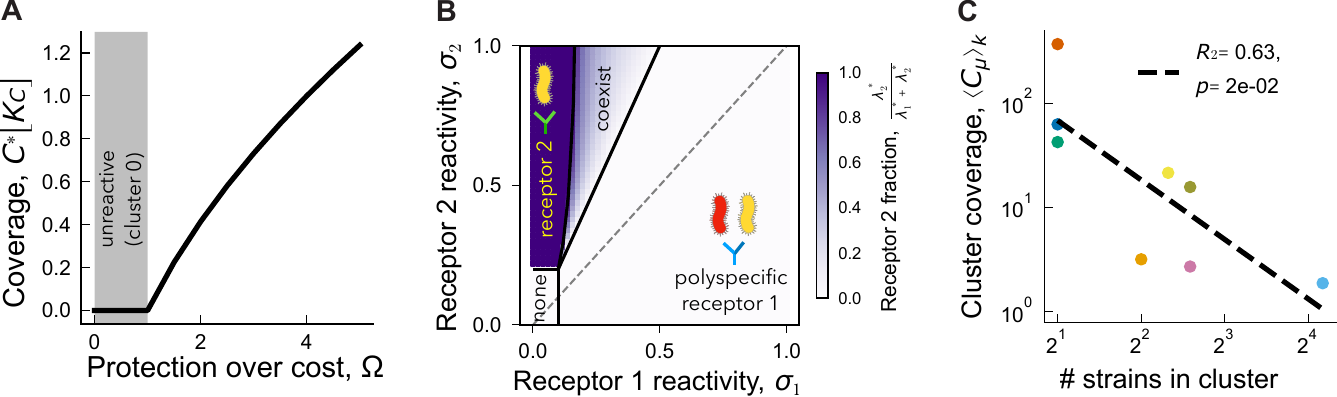}
\caption{{\bf Model of repertoire allocation explains repertoire sparsity, strain clustering and variation in immune coverage.} {\bf(A)} Optimal repertoire coverage ($C^*$) as a function of the marginal protection-to-cost ratio $\Omega$ (eq.~\ref{eq:one-type-optimal-immune-coverage}) in a repertoire where each receptor is reactive to a single strain. When this ratio falls below one, no receptor is deployed to target that strain. {\bf(B)} Phase diagram of receptor co-existence for two receptors ($r=1,2$) as a function of their reactivities for two strains ($\mu=1,2$). Receptor 1 is polyspecific, with equal reactivity to both strains ($\sigma_1 =\sigma_{\mu=1,r=1}=\sigma_{2,1}>0$), while receptor 2 is a monospecific receptor reactive to a single strain ($\sigma_{2} =\sigma_{2,2} >0;\, \sigma_{1,2}=0$). The heatmap shows the fraction of receptor 2 in the optimal repertoire, $\lambda_2^*/(\lambda_1^* + \lambda_2^*)$. The dashed gray line indicates the partition of reactivity space that would be expected between two monospecific receptors targeting the same strain. The region of viability for receptor 2 is highly restricted because receptor 1 tends to out-compete it in the optimal repertoire. Parameters: $h_1g_1^\text{max}/(K_C\beta) = h_2g_2^\text{max}/(K_C\beta) = 5$. {\bf(C)} Relationship between cluster repertoire coverage and cluster diversity in hCom1. Each colored marker represents a cluster (as in Fig.~\ref{fig3}A), and the dotted line shows a log-log fit between coverage ($\langle C_\mu\rangle_k$) and number of strains in cluster $k$ ($|k|$), $\log_{10}(\langle C_\mu \rangle_k) = m\log_{10}(|k|) + b$, in each cluster ($k=1,\dots,8$), with $m=-1.9$ and $b=2.4$.
}
\label{fig4}
\end{figure*}

Following eq.~\ref{eq:one-type-optimal-immune-coverage}, a receptor is allocated ($C^* >0$) if $\Omega >1 $ (Fig.~\ref{fig4}A); that is, if its conferred protection is greater than its cost. In the experiments we analyzed (Fig.~2), only 45 out of 97 strains were strongly reactive to the sampled T~cells and that this reactive subset did not include many high abundance strains in the microbiota~\cite{Nagashima2023-fe}. Similar observations have been made for immunoglobulin A (IgA) reactive strains in the gut~\cite{Palm2014-bz}. In the non-reactive group (cluster 0), three well-known human commensals account for nearly $50\%$ of total abundance: \textit{Bacteroides sp.}, \textit{Bacteroides thetaiotaomicron} and \textit{Akkermansia muciniphila}~\cite{Wu2011-do, Everard2013-fb}. With the caveat that all responding T~cells may not have been sampled, our model suggests that mounting T-cell-dependent responses to these commensal strains may not justify the associated costs in this microbiota.

Next, we asked if competitive exclusion between T~cell clones could explain the sparsity of the T~cell reactivities, where a few polyspecific T~cell clones cover taxonomically diverse strains. We modeled a system with two strains $\mu \in \{1,2\}$, and two T~cell clones with receptors $r \in \{1,2\}$, coupled by a fixed $2\times 2$ reactivity matrix $\Sigma$. Receptor 1 is polyspecific, with equal and unrestricted reactivity to both strains ($\sigma_{\mu=1,r=1}=\sigma_{2,1}=\sigma_1>0$) while receptor 2 is a monospecific receptor reactive to a single strain ($\sigma_{2,2} =\sigma_{2}>0;\, \sigma_{1,2}=0$). For each microbe--receptor pair ($\mu,r$), we define a protection-to-cost ratio, $\Omega_{\mu,r} = \sigma_{\mu,r} h_rg_\mu^\text{max}/(K_C\beta)$. As in the single receptor case, we find that whenever the reduction in harm from deploying either receptor does not exceed the marginal cost of immunity, $\sum_{\mu\in\{1,2\}}\Omega_{\mu,r}\leq1$ for $r \in \{1,2\}$, it is optimal to allocate no receptors to cover the strains: $\lambda_1 +\lambda_2 = 0$ (Fig.~\ref{fig4}B; see SI Appendix for derivation). 

In Fig.~\ref{fig4}B, we map out the co-existence regimes for the monospecific receptor. We find that for co-existence to be merely possible, the monospecific receptor must be more protective against its target (strain 2) than the polyspecific receptor: $\Omega_{2,2} > \Omega_{2,1}$ (SI Appendix). Coexistence is guaranteed only when the monospecific receptor (receptor 2) is sufficiently more protective than the polyspecific receptor:~$\Omega_{2,2} > \Omega_{2,1} + \Omega_{1,1}$. Alternatively, this condition requires the monospecific receptor to be appreciably more reactive than the polyspecific receptor:~$\sigma_2/\sigma_1 > 1 + h_1g_1^\text{max}/h_2g_2^\text{max}$. Thus, unless the strain targeted by the monospecific receptor (strain 2) is much more harmful than the other strain ($h_2g_2^\text{max} \gg h_1g_1^\text{max}$), the monospecific receptor can only co-exist with the polyspecific receptor in a small region of the phase diagram (Fig.~\ref{fig4}B; SI Appendix, Fig.~\SIfigfour). This region is smaller than would be expected between two monospecific receptors targeting the same strain (Fig.~\ref{fig4}B, gray dotted line), illustrating that a less reactive polyspecific receptor can exclude a more reactive monospecific receptor. Mathematical analysis of more complex microbiota-repertoire configurations recapitulates the finding that polyspecific receptors tend to exclude specialists, thereby biasing the repertoire towards a subset of minimally overlapping (i.e., sparse), broadly-reactive receptors (SI Appendix). This observation can be traced to the L1 regularizer in our objective function~(eq.~\ref{eq:cost_function}). 

Interestingly, the repertoire in the experiments we analyzed did not contain any receptors that bound to just one strain and there was little overlap in receptor reactivity patterns between clusters (Fig.~\ref{fig2}). Conversely, a recent experiment profiling the T~cell response to a 12-member microbial community in mice found a more balanced mix of specific and polyspecific clones~\cite{Vetter2026-lv}, suggesting that polyspecific clones are enriched in larger communities, where taxonomically unrelated strains are more likely to share antigens. This mirrors prior findings that broadly neutralizing antibodies, rather than antigen-specific ones, are favored when responding to diverse viral strains that share common antigens~\cite{Luo2015-gq, Nourmohammad2016-ed}.

When the optimal repertoire consists of a single polyspecific receptor $r^*$, whose reactivity to both strains is similar, $\sigma_{\mu,r^*} \approx \sigma >0$ for $\mu \in \{1,2\}$, the coverage of the strains is analogous to the single-strain case:

\begin{align}
C^*\approx\max\left [ \left (\sum_{\mu} \Omega_{\mu,r^*} \right) ^{1/2} -1 ,0 \right]\cdot K_C
\label{eq:rec-abund-2d},
\end{align}
where $\Omega_{\mu,r^*} = \sigma h_\mu g_\mu^\text{max}/(K_C\beta)$ is the dimensionless protection-to-cost ratio for each microbe-receptor pair. This expression generalizes straightforwardly to an arbitrary number of strains. Assuming comparable receptor reactivities $\sigma$ and microbial niche sizes (maximum abundance) $g^\text{max}$, we can ask how a microbe's harmfulness shapes immune coverage. A highly harmful microbe ($\Omega >1$) can on its own justify a dedicated receptor targeting it in the repertoire, as implied by eq.~\ref{eq:one-type-optimal-immune-coverage}. Less harmful strains, by contrast, can collectively justify a polyspecific receptor $r^*$ targeting their shared epitope whenever $\sum_\mu \Omega_{\mu,r^*} >1$, as implied by eq.~\ref{eq:rec-abund-2d}. This suggests that immune-reactive clusters containing many diverse strains may tend to contain less harmful strains, whereas small clusters are associated with more harmful strains. Indeed, in hCom1 the number of strains in a cluster is inversely related to the repertoire coverage averaged over the cluster's constituent strains (eq. \ref{eq:cluster_coverage}; Fig.~\ref{fig4}C).

Competitive exclusion has recently been proposed as a principle shaping optimal memory and regulatory T~cell repertoires~\cite{Mayer2015-ii, Marsland2021-ls}. Our model of repertoire allocation is mathematically equivalent to a biologically plausible model for consumer-resource competition between T~cell clones for limited microbe antigens (SI Appendix). Thus, it formalizes the biological intuition that the visibility of microbial antigens to the immune system, as described by the harmfulness constants ($h_\mu$), can be leveraged to rationally control the abundance of select microbes.

\subsection{T~cell repertoire coverage reflects cluster-level microbial harmfulness}

\begin{figure}[ht!]
\centering
\includegraphics[width=0.80\linewidth]{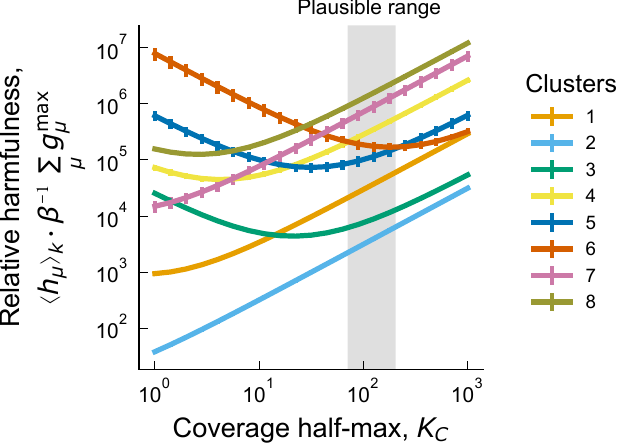}
\caption{{\bf Model of repertoire allocation infers relative harmfulness of strain clusters.} Relative harmfulness of strain clusters (eq.~\ref{eq:harm_estimate}) as a function of the half-maximal constant of immune inhibition of microbial abundance ($K_C$), and standard deviation (error bars) across 5 biological replicates for compositional data. The gray region denotes the $\pm 1.96$ standard deviation parameter range obtained by fitting eq.~\ref{eq:empirical_fit} to the data from Fig. \ref{fig3}C (SI Appendix, Fig.~\SIfigfive).
}
\label{fig5}
\end{figure}

We next sought to infer a measure of the relative harm of microbial clusters from their T~cell repertoire coverage. While the physiological roles of strains in complex communities are difficult to determine directly, the immune response provides a promising readout for distinguishing harmful from commensal strains~\cite{Palm2014-bz, Stein2018-dt, Vujkovic-Cvijin2022-wh, Probstel2020-pa, Olm2025-sy}. Assuming we are in the polyspecific-receptor regime, eq.~\ref{eq:rec-abund-2d} generalizes to a cluster of strains co-reactive to the same T~cell by summing over the protection-to-cost ratios of the strains. Inverting this generalized equation, we can relate the expected cluster coverage $C^*_k$ ($k = 1,\dots, 8$) to the composition-weighted (as in eq.~\ref{eq:cluster_coverage}) harmfulness of the cluster, $\langle h_\mu \rangle_k$: 

\begin{align}
\langle h_\mu\rangle_k(K_C) \approx \beta K_C\frac{\left(\langle C_\mu\rangle_k/K_C + 1\right)^2 }{\sigma_k\sum_{\mu ~\in~\text{cluster }k}g_\mu^\text{max}},
\label{eq:harm_estimate}
\end{align}
where optimal coverage is approximated by the observed average coverage over the strains in the same cluster: $C^*_k \approx \langle C_\mu \rangle_k$ (eq.~\ref{eq:cluster_coverage}; see SI Appendix for a detailed derivation).

To use eq.~\ref{eq:harm_estimate}, we approximate the reactivity of strains within the cluster ($\sigma_k$) as the composition-averaged maximum reactivity of strains within the cluster to all receptors $r$ in the repertoire, $\sigma_k \approx \langle \max_r\{\sigma_{\mu,r}\}\rangle_k$ (using data from Fig.~\ref{fig2}). The pre-coverage niche size of each cluster ($g_\mu^\text{max}$) in the denominator of eq.~\ref{eq:harm_estimate} is proportional to its relative abundance at week 1 (data from Fig.~\ref{fig3}A). This constant of proportionality $1/\sum_\mu g_\mu^\text{max}$ and the cost of immunity $\beta$ are unknown but identical across clusters, so harmfulness can be expressed in their combined units. Finally, we vary the remaining unknown parameter $K_C$ (eq.~\ref{eq:microbe-steady-noecology}), which determines how strongly coverage mitigates harm in our model (Fig. \ref{fig5}). At high $K_C$, harmfulness is mainly a function of abundance (abundance at week 1; Fig.~\ref{fig3}A), with the smaller clusters identified as more harmful. While at low $K_C$, harmfulness reflects both microbial abundance and repertoire coverage (Fig.~\ref{fig3}A,C), with harmfulness highest for highly-covered, low-abundance strains.

Although the precise value of $K_C$ is unknown, regulatory T~cells constitute $5-10\%$ of peripheral CD4+ T~cells in healthy individuals, suggesting a homeostatic, baseline immune coverage (eq.~\ref{eq:approx-immune-coverage}) $C^\text{low} \approx 10-20$~\cite{Sakaguchi2004-mc, Su2016-zo}. We expect the half-maximal constant $K_C$ to exceed this baseline level. Plugging in the steady-state equation of microbial abundance as a function of coverage (eq.~\ref{eq:microbe-steady-noecology}) into the centered log ratio (eq.~\ref{eq:cluster_clr}), we can derive a relationship between the compositional change and the observed immune coverage:

\begin{align}
\label{eq:empirical_fit}
    \Delta\text{CLR}_k =  -\log_{10}(1 + \langle C_\mu\rangle_k/K_C) + d,
\end{align}
where $d$ is a constant. We fitted this equation to the data in Fig.~\ref{fig3}C and obtained a plausible range for the half-maximal constant $K_C \approx 70.7-204$ ($\pm 1.96$ standard deviations); see Fig. \ref{fig5}; SI Appendix, Fig.~\SIfigfive. As expected, the inferred range is higher than known values of the homeostatic baseline. Within this plausible range for $K_C$, clusters 1 and 2 (commensals) are robustly inferred as lower-harm, and cluster 6 (pathobionts) as higher-harm across a broad range of $K_C$ (Fig. \ref{fig5})~\cite{Henke2019-ry, Robert2007-ig, Elsayed2004-lv}. Experiments that simultaneously titrate coverage and measure abundance changes could pin down $K_C$ and tighten these inferences~\cite{Olm2025-sy}.

\subsection{Microbial ecology can limit the efficacy of immune coverage}

\begin{figure*}[ht!]
\centering
\includegraphics[width=0.8\linewidth]{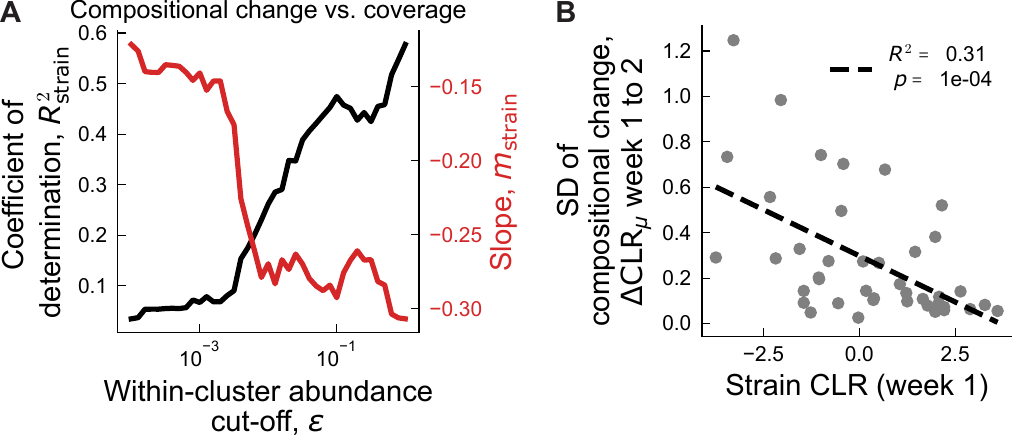}
\caption[Microbial ecology can modulate the efficacy of immune coverage]{{\bf Microbial ecology can modulate the precision of immune control.} {\bf(A)} Regression analysis between the strain-level microbial abundance change $\Delta\text{CLR}_\mu$ and immune coverage $C_\mu$, modeled as $\Delta\text{CLR}_\mu = m\log_{10}(C_\mu) + b$; the coefficient of determination $R^2_\text{strain}$ (left) and the regression slope $m_\text{strain}$ (right) are reported. Regressions are performed on strain subsets defined by an abundance cut-off $\epsilon$ (horizontal axis): a strain is excluded if its week-1 abundance falls below $\epsilon$ times that of the most abundant strain in its cluster. Larger $\epsilon$ therefore corresponds to more stringent filtering, retaining only the highest-abundance strains within each cluster. {\bf(B)} Standard deviation (SD) of individual strain composition changes ($\Delta\text{CLR}_\mu$) from week 1 to 2 versus mean across replicates ($n = 5$; gray dots). Data was fit to: $\text{SD} = m\cdot\text{CLR} + b$. The inverse relationship (black line; $m =-0.08,~b =0.3$) points to greater measurement noise in less abundant strains than in more abundant ones.
}
\label{fig6}
\end{figure*}

A puzzling feature of the data is that the strong inverse relationship observed between cluster-level compositional changes and immune repertoire coverage ($R^2_\text{cluster} = 0.66$; Fig.~\ref{fig3}C) all but disappears when these clusters are disaggregated into their constituent strains ($R^2_\text{strain}=0.02$; Fig.~\ref{fig3}D). Repertoire coverage thus predicted the abundance dynamics of strain clusters (and their targeted antigens), but failed to capture the trajectories of individual strains within them. Strikingly, this strain-level discrepancy was concentrated among low-abundance strains within each cluster. Progressively filtering out the lowest-abundance strains from each cluster restored the coverage--abundance relationship at the strain level (Fig.~\ref{fig6}A). 

As this low-abundance discrepancy might simply reflect stochastic measurement noise, we asked if measurements of compositional change were noisier for less abundant strains than more abundant strains. We found a robust inverse relationship between noise magnitude and abundance (Fig.~\ref{fig6}B); however, averaging out this noise failed to recover the cluster-level relationship between immune coverage and compositional change ($R^2 = 0.04,~m=-0.01$; SI Appendix, Fig.~\SIfigsix). This suggests that strain-level dynamics and not measurement noise might better explain the discrepancy.

Ecological interactions among strains provide a possible mechanism for the discrepancy. We reasoned that if there are ecological interactions between strains in a cluster, then less abundant strains might benefit when repertoire coverage decreases the abundance of more dominant strains within their cluster, akin to ``competitive release''~\cite{Segre2016-ak, Newton2019-xv, Grilli2017-tg, Angulo2019-md, Newton2023-ln}. To formalize this, we modeled the steady-state abundances of a minimal two-strain system under shared immune coverage ($C$), where a dominant strain 1 suppresses a subdominant strain 2 with interaction strength $s> 0$:

\begin{align}
\tilde{g}_1(C) = \frac{g_1^\text{max}}{1 + C},~~
\tilde{g}_2(C) =\frac{g_2^\text{max}}{1 + C + s\tilde{g}_1(C)},
\label{eq:eco_interactions}
\end{align}
where $g^\text{max}_{1/2}$ are their respective carrying capacities. For any combination of immune coverage ($C$) and ecological interaction ($s$), the composition change gap (from $C=0$ to $C> 0$) between the two strains is given by (SI Appendix):

\begin{align}
\Delta\text{CLR}_1 - \Delta\text{CLR}_2= -\log_{10}\left(\frac{1 + sg^\text{max}_1}{1 + sg^\text{max}_1/(1 + C)^2} \right).
\label{eq:eco_gap}
\end{align}
For competitive interactions ($s>0$), eq.~\ref{eq:eco_gap} predicts a composition change gap that grows increasingly negative as immune coverage ($C$) increases. Intuitively, as coverage drives down the dominant strain~1, the competitive pressure it exerts on strain~2 is relieved. Strain~2 therefore declines less than strain~1, widening the gap between them. This result does not depend on the exact model details and can be obtained with an alternate model for ecological competition (SI Appendix). Finally, we observed composition change gaps between dominant and sub-dominant strains across microbial clusters in the data that cannot be explained by differences in strain-specific immune coverage (SI Appendix, Figs.~\SIfigseven, ~\SIfigeight). Ecological coupling may thus convert immune suppression of a dominant strain into a relative gain for its subdominant cluster-mate.
 
\section{Discussion}

A central difficulty in microbiome research is that the taxonomic composition of the gut community varies widely between healthy hosts, even as several host-relevant outputs---such as stable metabolite production and the reproducible antibody-coating of microbial strains---remain comparatively conserved~\cite{Arnoldini2025-my, Palm2014-bz}. This contrast suggests that simple, robust mappings can link a complex community to host-relevant properties, yet we still lack principled ways to derive them. Here we provide one. Drawing on recent experiments that measured immune interactions with strains in germ-free mice inoculated with a defined microbial community~\cite{Nagashima2023-fe, Cheng2022-mv, Wang2023-tc}, we developed mathematical models that link immune reactivity to microbial dynamics. Our model treats the gut-associated CD4$^+$ T cell repertoire as an informative proxy of how inhibitory pressure is allocated to minimize microbe-driven systemic harm. This framework suggests that the repertoire operates not on individual species but on their shared molecular antigens, collapsing an ecologically complex community onto a small number of clusters with distinct immune reactivities that are largely orthogonal to taxonomy.

A striking feature of the gut immune--microbe interaction data is its sparsity: a small set of polyspecific T~cell receptors recognizes antigens shared across taxonomically diverse strains, with no responding receptor binding a single microbe and little overlap in reactivity between clusters. This sparsity could simply reflect a limited availability of cognate receptors or microbial antigens; two lines of evidence argue otherwise. First, the peripheral repertoire cycles through the gut epithelium, making a diverse, organism-wide receptor pool available for microbe recognition~\cite{Lockhart2024-tq}. In mice this pool contains approximately $10^6$ unique clones, orders of magnitude more than the handful of responding clones identified in the data we analyzed~\cite{Moon2007-pk}. Second, despite the structural complexity and multi-antigen nature of taxonomically diverse strains~\cite{Sonnert2024-op}, the T~cell response to a systemic infection is typically clonally diverse and can target multiple antigens derived from a single pathogen~\cite{Parsa2022-js, Parsa2026-ko}. The focused targeting of specific clusters thus reflects a rational allocation strategy rather than happenstance. Our optimization framework makes this explicit. A receptor is recruited only when its protective benefit outweighs the cost of the immune response, and the generic optimum is a sparse set of polyspecific receptors recognizing shared antigens, consistent with the data. Notably, this optimal allocation maps onto the steady state of consumer--resource competition among T cell clones for limited antigens~\cite{Mayer2015-ii, Marsland2021-ls}, and so need not be computed by a central controller but can emerge dynamically from inter-clonal competition.

Immune coverage further functions as a quantitative readout of microbial harm. Coverage strongly predicts cluster-level compositional dynamics, distinguishing commensal-dominated clusters from pathobiont-containing clusters, the latter of which are suppressed but not eliminated from the microbial ecology. By attributing coverage to the harm associated with a cluster, our model yields a relative-harmfulness ranking of the microbial clusters in the data. One of the most harmful clusters contains strains of \textit{Ruminococcus gnavus} which has been associated with Crohn's disease~\cite{Henke2019-ry}, consistent with its inferred relative harmfulness. Yet we caution that the inferred harmfulness is a composite quantity that cannot be attributed to a single mechanism and is best interpreted as one of several factors, alongside its abundance, that influence a strain's visibility to the immune system.

The strong coverage--composition relationship at the cluster level vanishes at the level of individual strains, particularly the low-abundance ones. We showed that sampling noise in detecting rare microbes cannot account for this on its own. The discrepancy could be in part due to spatial compartmentalization in the gut \cite{Welch2017-vi}. Individual strains might escape host immune pressure if they express a targeted antigen away from the site of an immune response. Rigorously evaluating this hypothesis requires pairing experiments that collect spatial data on microbial composition with direct measurements of interactions between antigen presenting cells and T~cells, such as LIPSTIC labeling~\cite{Nakandakari-Higa2024-gb}. Another non-exclusive explanation is ecological coupling and competitive release. In a minimal model of ecological interactions we showed that as immune coverage drives down the dominant strain in a cluster, the competitive pressure it exerts is relieved, allowing less abundant strains to rise in relative abundance despite the immune pressure they experience. The strain pair (\textit{R. gnavus} and \textit{C. scindens}) in cluster 6 that experiences the strongest immune coverage (Fig.~\ref{fig2}) offers an ideal test of this hypothesis. Co-culture, selective depletion of the dominant partner, or pairwise reconstitution in germ-free hosts would test whether subdominant strain dynamics are strongly driven by their dominant partner, as the model predicts. Such experiments can be used to decouple ecological and immune forces shaping the microbiota.

Beyond explaining the present data, our model can offer design principles for minimally disruptive microbiota therapies. Altering a strain's ecological niche shifts its protection-to-cost ratio, which can in turn reorganize the associated immune repertoire, and vice versa~\cite{Littmann2021-om}. This implies that compatibility for interventions such as fecal matter transplantation may not require precise strain-level matching; instead, aligning the broad antigen compositions of donor and host may suffice.

Several limitations bound these conclusions. Our study relies on a modest sample size ($n=5$ mice for compositional data; $n=3$ mice for pooled T~cell repertoires), which precludes independent verification of the universality of these specific microbial clusters across identically inoculated hosts. Additionally, our assumption that the host-microbiota interaction actively optimizes repertoire allocation requires mechanistic validation. Nevertheless, our results demonstrate that data-driven mathematical approaches, which have been successful in other contexts~\cite{Mayer2015-ii, Angulo2019-md, Marsland2021-ls, Schnaack2021, Good2023-cv, Ho2024-oz, Feng2025-ml, Ukogu2026-ag}, can rationalize complex immune--microbiota interactions. Future studies pairing longitudinal, single-animal microbiota profiling with single-cell TCR sequencing, metabolic outputs, and spatial tracking tools will be essential~\cite{Nakandakari-Higa2024-gb, Olm2025-sy, Furuichi2024-xf, Wang2023-tc, Belkaid2014-ay}. These should ideally be complemented by control studies in immunocompromised (e.g., $\textit{Rag}^{-/-}$) mice to isolate purely ecological dynamics~\cite{Littmann2021-om, Ameline2025-xm, Zhang2015-xc}, alongside targeted strain-invasion or depletion experiments. Together, these efforts will be essential for determining how the immune repertoire reorganizes in response to a complex microbial ecology in a changing environment.

\section{Materials and Methods}
\noindent Data and code for all the analyses can be found in the Github
repository:~\url{https://github.com/obinna-ukogu/gutimmune}. Previously published data from refs.~\cite{Cheng2022-mv, Wang2023-tc,Nagashima2023-fe} were used for this work.

\section{Acknowledgment}
This work has been supported by the CAREER award from the National Science Foundation grant 2045054 (A.N., O.U.), Agence Nationale de la Recherche grant ANR-24-CE45-7957 ``WILDTYPES'' grant (T.M., A.M.W.) and the EvoMG HORIZON-MSCA-2024-DN-01-01 (T.M., A.M.W.). This work is also supported, in part, through the Departments of Physics and Applied Mathematics and the College of Arts and Sciences at the University of Washington. This work benefited from discussions during the 2024 program ``Interactions and Co-evolution between Viruses and Immune Systems'' at the Kavli Institute for theoretical physics, which is supported by the National Science Foundation grant PHY-2309135, and the Gordon and Betty Moore Foundation Grant No. 2919.02.

\bibliographystyle{plos2015}
\bibliography{references}

\end{document}


\noindent{\bf \Large Supplementary Information\\}
\vspace{-1ex}

\noindent {\bf \large Rational T~cell repertoire allocation to control the gut microbiota}\\

\vspace{-2ex}
\noindent Obinna A. Ukogu, Angelina M. Bilate, Thierry Mora, Aleksandra M. Walczak \& Armita Nourmohammad

\tableofcontents

\clearpage

\section{Data analysis}

\subsection{Clustering of strains}
Reactive strains were grouped using $K$-means clustering based on their $L$-1 normalized (column) reactivities to 15 responsive T cell clones (Fig.~\mainfigtwo). We selected the cluster number $K = 8$ that nearly maximized the mean silhouette score (implemented in Python \cite{scikit-learn}), 
\[
\sum_{\mu=1}^{45}[b(\mu) - a(\mu)]/ \max[a(\mu),b(\mu)],
\] 
where $a(\mu)$ is the mean distance between strain $\mu$'s reactivity vector and every other point in its assigned cluster, and $b(\mu)$ is the mean distance between strain $\mu$'s reactivity vector and every other point in its nearest, neighboring cluster. The mean silhouette coefficient measures the cohesiveness of clusters and how distinct they are from each other.

\subsection{Processing and transformation of microbial composition data}

Before analysis, strains that were undetected across all replicates at week 1 were dropped from the analysis. Otherwise, missing relative abundance data was set to $10^{-8}$, consistent with the reported limit of detection in the study \cite{Cheng2022-mv}. Relative abundance data for all strains were then normalized to sum to 1, per time-point and replicate.

To quantify compositional shifts at the strain level, we computed the centered log-ratio (CLR) transform of strain relative abundances at each time $t > 0$, with $g_\mu(t)\Big/\sum_{m =1}^Mg_m(t), ~\mu=1,\dots,M$. For each strain, the CLR is
\begin{align}
\begin{split}
 \text{CLR}_\mu{(t)} = \log_{10}\left(\frac{g_\mu(t)}{\sum_{m =1}^Mg_m(t)}\Bigg/\sqrt[M]{\prod_{\alpha = 1}^{M} \frac{g_\alpha(t)}{\sum_{m =1}^Mg_m(t)}}~\right),\\
\end{split}
\label{eq:strain_clr}
\end{align}
where $M = 97$ and the total abundance ($\sum_{m =1}^Mg_m(t)$) cancels out of the numerator and denominator of the expression inside the logarithm. This transformation minimizes spurious correlations introduced by the sum-to-one constraint of relative abundance data of the $M$ microbial strains. We can then compute the compositional change of each strain $\mu$ from time point $t_1$ to $t_2$:
\begin{align}
\begin{split}
\Delta\text{CLR}_\mu(t_1,t_2) &= \text{CLR}_\mu(t_2) - \text{CLR}_\mu(t_1)\\
&= \log_{10}\left(\frac{g_\mu(t_2)}{g_\mu(t_1)} \right) - \frac{1}{M}\sum_m^M\log_{10}\left(\frac{g_m(t_2)}{g_m(t_1)}\right),
\end{split}
\label{eq:delta_strain_clr}
\end{align}
from which we observe that the first term is related to the abundance fold-change of the strain $g_\mu(t_2)/g_\mu(t_1)$, and the second term depends on the fold-change of every strain in the microbiota. If we consider the gap in the change in CLR between two strains $(\mu, \alpha)$,
\begin{align}
\Delta\text{CLR}_\mu(t_1,t_2) - \Delta\text{CLR}_\alpha(t_1,t_2) = \log_{10}\left(\frac{g_\mu(t_2)}{g_\mu(t_1)} \right) - \log_{10}\left(\frac{g_\alpha(t_2)}{g_\alpha(t_1)} \right),
\end{align}
this quantity depends only on the abundance fold-change of the two strains. Analogous expressions for the CLR of microbial clusters are given in the main text.

Because strains within a cluster map to different measured immune coverages (SI Fig.~\mainfigtwo), their abundance trajectories are expected to differ. To better characterize the intra-cluster differences in the impact of immune coverage on microbial abundances, we quantified the composition change gap ($\Delta\text{CLR}_{\text{1st}^{(k)}}-\Delta\text{CLR}_{\text{rest}^{(k)}}$) between the most abundant strain (``1st'') and the remaining lower-abundance members (``rest'') of each cluster ($k=1,\dots,8$ indicates cluster identity; SI Fig.~\ref{SIfig7}). 

We constructed a model based on the empirical fit between the compositional change of the most abundant strain in each cluster and its immune coverage (Fig.~\mainfigsix A). This fit is defined as: $\Delta\text{CLR} = m_\text{strain}^{(\epsilon = 1)}\log_{10}(C) + b$. First, we calculated the expected gap by assuming the measured immune coverage in SI Fig.~\mainfigtwo~reflected the true interaction between each strain and the repertoire, with
\[
\Delta\text{CLR}_{\text{1st}^{(k)}}-\Delta\text{CLR}_{\text{rest}^{(k)}}= m_\text{strain}^{(\epsilon=1)}\log_{10}\left(\frac{C_{\text{1st}^{(k)}}}{\langle C_{\mu \neq \text{1st}^{(k)}}\rangle_k}\right), 
\]
where $\langle C_{\mu \neq \text{1st}^{(k)}}\rangle_k$ is the strain-abundance-weighted average of measured repertoire coverage of the less abundant strains in each cluster. 
A difference in coverage thus translates into a gap in compositional change between most abundant strain and the remaining strains (weighted by their week 1 abundances) in each cluster. Intriguingly, the observed gap was frequently negative, deviating from both models and exceeding the expected gap for highly covered clusters (SI Fig.~\ref{SIfig7}). This effect was most pronounced in cluster 6, which contains two strains specialized in arginine metabolism~\cite{Wang2023-tc}. Within this cluster, the dominant pathobiont \textit{Ruminococcus gnavus} sharply declined relative to the entire microbiota while its less abundant cluster-mate \textit{Clostridium scindens} simultaneously expanded (SI Fig.~\ref{SIfig8}). Conversely, this trend was absent in less immunogenic, more taxonomically diverse commensal clusters 1, 2, and 7. These results demonstrate that the strain-level breakdown is consistent with a systematically biased gap, wherein less abundant strains benefit as their dominant cluster-mates are suppressed by host immunity.




\section{Model of optimal repertoire allocation}

\subsection{Model definition and analysis}

Consider a system with $M$ strains and $R$ receptors, and immune coverage
\begin{align}
\label{eq_gutimm: immune-coverage}
 C_\mu = \sum_{r \in \{1,\dots, R\}} \sigma_{\mu,r}\lambda_r,
\end{align}
for $\mu = 1,\dots, M$, where $\lambda_r$ corresponds to an inflammatory immune receptor/antibody $r$. For concreteness, we take
\begin{align}
\label{eq_gutimm: strain-steady-noecology}
\begin{split}
\tilde{g}_\mu(C_\mu) &= \frac{g_\mu^\text{max}}{1 + C_\mu/K_C},
\end{split}
\end{align}
where $K_C$ is the half-maximal constant of inhibition of strain abundance. In the analyses that follow, we make the convenient change of variables, $K_C^{-1}C_\mu \mapsto C_\mu$, expressing repertoire coverage in units of the unknown half-maximal constant of inhibition.

We assume that once in the gut a strain causes harm to the host organism at rate proportional to $h_\mu$, and further that the objective of the immune response is to control the steady-state population of strains in the gut to minimize harm. Given a set of available receptors $\{r=1,\dots,R~|~\Sigma\}$, we seek the solution to the optimization problem
\begin{align}
\label{eq_gutimm:cost_function}
\min_{\vec\lambda \geq 0}\left\{ F(\vec\lambda;\Sigma) \right\} = \min_{\vec\lambda \geq 0}\left\{ \sum_\mu h_\mu \tilde{g}_\mu(\vec\lambda) + I(\vec\lambda) \right\},
\end{align}
where the first term is the steady-state rate of harm induced by the strain and the second term corresponds to additional costs from maintaining an immune repertoire. Here, we specify the linear objective function,
\begin{align}
I(\lambda) = \sum_r \beta\lambda_r.
\end{align}
For example, the costs may describe an increased risk of autoimmunity associated with having more receptors, and/or the difficulty of generating particular receptors, and/or competition among receptor-bearing cells for limited survival signals. With $\tilde{g}_\mu(\cdot)$ defined as in eq. \ref{eq_gutimm: strain-steady-noecology}, we will solve
\begin{align}
\label{eq_gutimm: one-type-objective-func}
\min_{\lambda_r \geq 0}\left\{ \sum_{\mu =1}^M h_\mu \frac{g_\mu^\text{max}}{1 + C_\mu(\vec\lambda)} + \sum_{r = 1}^R\beta \lambda_r \right\}.
\end{align}

\subsubsection{Optimality conditions} 
To find the optimal solution, we write down the Lagrangian:
 \begin{align}
\label{eq_gutimm: general_lagrange}
 L_\beta(\vec\lambda, \boldsymbol{\delta}) = \sum_\mu h_\mu \frac{g_\mu^\text{max}}{1 + C_\mu(\vec\lambda)} + \sum_r\left[\beta \lambda_r + (l_r^2 -\lambda_r)\delta_r\right],
\end{align}
where we have introduced slack variables, $l_r \in \mathbb{R}$, and Lagrange multipliers, $\delta_r \in \mathbb{R}$, $r =1,\dots,R$, to account for the non-negativity constraints on $\vec\lambda$. Then, 
\begin{align}
\label{eq_gutimm: one-type-first-derivative}
 \begin{split}
  \frac{\partial L_\beta}{\partial \lambda_r} &= \sum_\mu h_\mu\frac{\partial C_\mu}{\partial \lambda_r}\frac{\partial \tilde{g}_\mu}{\partial C_\mu}(\vec\lambda) + \beta -\delta_r,\\
  \frac{\partial L_\beta}{\partial \delta_r} &= l_r^2 -\lambda_r,
 \end{split}
\end{align}
for $r = 1,\dots, R$, where
\begin{align}
 \frac{\partial C_\mu}{\partial \lambda_r}=\sigma_{\mu,r}.
\end{align}
so that, following the KKT conditions, solutions satisfy
\begin{align}
\label{eq_gutimm: one-type-KKT}
 \begin{split}
  \frac{\partial L_\beta}{\partial \lambda_r} &= \sum_\mu \sigma_{\mu,r}h_\mu\frac{\partial \tilde{g}_\mu}{\partial C_\mu}(\vec\lambda) + \beta -\delta_r = 0,\\
  \frac{\partial L_\beta}{\partial \delta_r} &= l_r^2 -\lambda_r = 0,
 \end{split}
\end{align}
for $r = 1,\dots, R$. If $\vec\lambda^*$ is a local minimizer, then $\delta_r^* \geq 0, ~r=1,\dots,R$, and if $\delta_j^* > 0$ for some $j \in \{1,\dots,R\}$, then we must be on a boundary, $\lambda_j^* = 0$. Define the rate of change in harm from killing strain $\mu$ as
\begin{align}
\label{eq_gutimm: one-type-marginal}
\begin{split}
 y_\mu(\vec\lambda) = h_\mu\frac{\partial \tilde{g}_\mu}{\partial C_\mu}(\vec\lambda)= -\frac{g_\mu^\text{max} h_\mu}{(1+C_\mu(\vec\lambda))^2}
\end{split}
\end{align}
where the sign of $y_{\mu}$ is determined by $h_\mu$, the harmfulness of strain $\mu$, and
\begin{align}
\label{eq_gutimm: one-type-bound}
\begin{split}
 |y_\mu|< \left|g_\mu^\text{max} h_\mu\right|.
\end{split}
\end{align}
We will focus on the first system of equations which describes the extremal conditions for $\lambda_r$. The second set of equations imposes the non-negativity constraint on $\lambda_r$, and we will check after we have identified the potential minima.

\subsubsection{Solving the general problem} 
We seek the most protective immune coverage of the microbiota with the fewest number of immune cells (and receptors). The nature and existence of solutions depends strongly on the reactivity matrix $\Sigma$. The KKT conditions require
\begin{align}
\label{eq_gutimm: linear system}
\delta_r -\beta= \sum_\mu h_\mu\frac{\partial C_\mu}{\partial \lambda_r} \frac{\partial \tilde{g}_\mu}{\partial C_\mu} = \sum_\mu \sigma_{\mu,r} y_\mu \Longrightarrow \Sigma^\top \boldsymbol{y} =\boldsymbol{\delta}-\beta\boldsymbol{1}.
\end{align}
If $\vec\lambda^*$ is a local minimizer, then $\delta_r^* \geq 0, ~r=1,\dots,R$, and if $\delta_j^* > 0$ for some $j \in \{1,\dots,R\}$, then we must be on a boundary, $\lambda_j^* = 0$.\footnote{From eq. \ref{eq_gutimm: one-type-KKT}, we can infer that $\delta_r^* \leq \beta$.} This tells us immediately that if $\delta_r^* < 0$ at the boundary then $\vec\lambda^*$ is not a local minimum, since it means it is possible to minimize the objective further by increasing $\lambda_r^* =0 \to \lambda_r^* > 0$. Computationally, finding the optimal solution boils down, in part, to selecting which $R^* \leq R$ receptors to use ($\delta_r^* = 0, ~r \in \kappa_+ \subseteq\{1,\dots,R\}$), and which not to use ($\delta_j^* > 0,~\lambda_j^* = 0, ~ j \in \{1,\dots,R\}\setminus\kappa_+$), and solving the system of equations for a $\boldsymbol{y}^*$ that is in the functional range of eq.~\ref{eq_gutimm: one-type-marginal},
\begin{align}
\label{eq_gutimm: one-type-reduced_system}
\begin{split}
\tilde{\Sigma}^\top \boldsymbol{y}({\vec\lambda^*}) &= -\beta\tilde{\boldsymbol{1}} ~~\text{for} ~r \in \kappa^{+}, \\ \delta_j^* &> 0 ~~\text{for}~ j \in \{1,\dots,R\}\setminus\kappa_+ , ~ ~~\text{(Local minimizer)}
\end{split}
\end{align}
where $\tilde \Sigma^\top$ and $\tilde{\boldsymbol{1}}$ only include $R^*$ rows (corresponding to receptors) in $\kappa_+$.\footnote{If the rows of $\tilde \Sigma^\top$ are orthogonal, then the solution takes the form $\vec y^* = \sum_{r \in \kappa_+} \beta_r \frac{\tilde \Sigma_r}{||\tilde \Sigma_r||_2^2}$, where $\Sigma_r$ is the $r$-th column of $\tilde \Sigma$. Since the entries of $\tilde\Sigma$ are non-negative, this is only possible if each receptor covers a distinct group of strains.} Naively, there are $2^R$ possible receptor configurations. 

Generically the system is overdetermined unless $|\kappa_+| = R^* \leq M$, hence at most $M$ receptors are utilized. However, additional receptors may be able to co-exist if there exists at least one receptor $j$ whose reactivity vector $\sigma_j$ satisfies a strict condition:
\begin{align}
\label{eq_gutimm: general_coexistence}
\sigma_j = \sum_{r \neq j} c_r\sigma_r \geq 0,~~~\text{with}~ \sum_{r\neq j} c_r = 1,
\end{align}
that is, the reactivities of receptor $j$ is an affine combination of some other receptors that coexist in the repertoire. Violations lead to species extinctions. A receptor $j$ such that $\sum_{r\neq j} c_r < 1$ cannot be in the optimized repertoire since
\begin{align}
\delta_j = -\sum_\mu \sigma_{\mu,j} |y_\mu| + \beta= -\sum_\mu \sum_{r \neq j} c_r\sigma_{\mu,r} |y_\mu| + \beta= -\sum_{r \neq j} c_r\sum_\mu \sigma_{\mu,r} |y_\mu| + \beta= -\beta\sum_{r \neq j} c_r  + \beta > 0,
\end{align}
which means that it would be possible to minimize the objective by decreasing the abundance of receptor $j$, driving it to extinction. Conversely, if $\sum_{r\neq j} c_r > 1$ then $\delta_j < 0$, receptor $j$ will be recruited, displacing at least one other receptor from the repertoire. In either case, the reactivity matrix can be modified (eq.~\ref{eq_gutimm: one-type-reduced_system}), $\Sigma \to \tilde\Sigma$ to exclude extinct species. Eq.~\ref{eq_gutimm: general_coexistence} is a stringent condition on the entries of $\Sigma$ (and its $2^R$ possible subsets of co-existing receptors), so matrices satisfying it occupy a vanishingly small fraction of the space of possible reactivity matrices. We therefore expect $R^* \leq M$ for essentially any reactivity matrix.

If a vector $\boldsymbol{y}^* < 0$ can be found that solves the the reduced system of linear equations (eq. \ref{eq_gutimm: one-type-reduced_system}), we check that it is a feasible solution for the original non-linear problem and that the solution satisfies eq.~\ref{eq_gutimm: one-type-bound}. This requires solving the non-linear equation for $\vec{C}^* \geq 0$ (eq. \ref{eq_gutimm: one-type-marginal}) to find
\begin{align}
\label{eq_gutimm: one-type-optimal-immune-coverage}
 C_\mu^* =\left[\frac{g_\mu^\text{max} h_\mu}{|y_\mu^*| }\right]^\frac{1}{2} - 1.~~\text{(Optimal immune coverage)}
\end{align}
If $\vec{C}\geq 0$, to find $\vec\lambda^*$ we need to solve the linear system of equations (eq.~\ref{eq_gutimm: immune-coverage}),
\[
 C_\mu^* = \sum_{r \in \{1,\dots, R\}} \sigma_{\mu,r}\lambda_r^*
\]
for $\mu \in \{1,\dots,M\},$ with $\lambda_j = 0$ if $j \in \{1,\dots,R\}\setminus\kappa_+$ 
(i.e. $\delta_j > 0$). Finally, we must verify that the solution we found is non-negative $\vec\lambda^* \geq 0$.



\subsubsection{Convexity of the objective function} 
As in the previous discussion, the convexity of the objective function depends strongly on the reactivity matrix $\Sigma$. Naively, computing the second derivative for a general objective function $L_\beta(\vec{C}(\vec\lambda))$ satisfying $\frac{\partial^2 L_\beta}{\partial C_\mu^2}(\vec{C}(\vec\lambda)) > 0$ for $\mu \in \{1,\dots,M\}$:
\begin{align}
\begin{split}
 H_{j,r} = \frac{\partial^2 L_\beta}{\partial\lambda_r\partial\lambda_j} &= \sum_\mu \sigma_{\mu,j}\sigma_{\mu,r}\frac{\partial^2 L_\beta}{\partial C_\mu^2}\\
 &=\sum_\mu \sigma_{\mu,j}\left[\frac{\partial^2 L_\beta}{\partial C_\mu^2}\right]^\frac{1}{2}\times \sigma_{\mu,r}\left[\frac{\partial^2 L_\beta}{\partial C_\mu^2}\right]^\frac{1}{2}\\
 &= (A_j)^\top A_r
\end{split}
\end{align}
where the $r$-th column of $A$ is
\[
A_r^\top = \left[\left(\frac{\partial^2 L_\beta}{\partial C_\mu^2}\right)^\frac{1}{2}\sigma_{\mu,r}\right]_{\mu = 1}^{M}
\]
and 
\begin{align}
\begin{split}
A &= \text{diag}\left( \left[\left(\frac{\partial^2 L_\beta}{\partial C_\mu^2}\right)^\frac{1}{2}\right]_{\mu = 1}^{M} \right)\times\Sigma \\
H & = A^\top A.
\end{split}
\end{align}
Then $H$ is positive semi-definite, since
\begin{align}
\label{convexity-of-objective}
v^\top H v = v^\top A^\top Av = (vA)^\top Av \geq 0,
\end{align}
which is a quadratic form. This shows that the function is convex, although it is not strictly convex unless $\Sigma$ is full rank. However, as we have previously seen, modifying the reactivity matrix to remove unutilized receptors will generally lead to a full rank matrix and restore strict convexity (eqs.~\ref{eq_gutimm: one-type-reduced_system},~\ref{eq_gutimm: general_coexistence}).

\subsection{Ecological competition for antigens optimizes the repertoire}
Following \cite{Mayer2015-ii}, we ask if competition for antigens can generate the optimal repertoire. Specifically, we claim the ecological dynamics
 \begin{align}
\label{eq_gutimm: one-type-receptor-dynamics}
 \begin{split}
  \frac{\mathrm{d} \lambda_r}{\mathrm{d} t} &= \lambda_r\left(-\sum_\mu \sigma_{\mu,r}h_\mu\frac{\partial \tilde{g}_\mu}{\partial C_\mu}(\vec\lambda) - \beta\right),
 \end{split}
\end{align}
achieves this modified gradient descent of the objective function via competitive dynamics, where multiplying by $\lambda_r$ ensures that its value is never negative and $\beta$ acts like a death rate. The term
\[
h_\mu\frac{\partial \tilde{g}_\mu}{\partial C_\mu}(\vec\lambda) \to 0, ~\text{as}~ g_\mu \to 0 ~\text{or}~ h_\mu \to 0,
\]
corresponds to the availability of strain-derived antigens. With strain-immune response functions as in eq. \ref{eq_gutimm: strain-steady-noecology}, we have
\[
-\frac{\partial \tilde{g}_\mu}{\partial C_\mu}(\vec\lambda) = \frac{g_\mu^\text{max}}{(1+C_\mu)^2}
\]
which models a monotonically decreasing antigen supply as immune coverage increases.

Since any point that satisfies the KKT conditions (eq. \ref{eq_gutimm: one-type-KKT}) is a fixed point for the ecological dynamics, $d\lambda_r/dt = 0, ~r = 1,\dots,R,$ it is immediate that the optimal repertoire solution is a fixed point of the dynamics. Next, we show that stable fixed points of the dynamics are exactly the solutions of the KKT equations. For a fixed point to be stable the Jacobian matrix computed at the point must have only negative eigenvalues. The Jacobian is given by
\begin{align}
 J_{r,r'} = \delta_{r,r'}\cdot\left(-\sum_\mu \sigma_{\mu,r}h_\mu\frac{\partial \tilde{g}_\mu}{\partial C_\mu}(\vec\lambda) - \beta\right) - \lambda_r\sum_\mu\sigma_{\mu,r}\sigma_{\mu,r'}h_\mu\frac{2g_\mu^\text{max}}{(1 + C_\mu)^3}.
\end{align}

When $\lambda_r = 0$, row $r$ will have non-zero entries everywhere except the diagonal. The requirement that the eigenvalue of the elementary vector $\vec e_r$ (with 1 at position $r$ and zero elsewhere) be negative means that
\[
\sum_\mu \sigma_{\mu,r}h_\mu\frac{\partial \tilde{g}_\mu}{\partial C_\mu}(\vec\lambda) + \beta >0,
\]
which is equivalent to the KKT conditions at the boundary (eq. \ref{eq_gutimm: one-type-KKT}). When $\lambda_r > 0$ (and $\dot{\lambda}_r = 0$),
\begin{align}
\begin{split}
J_{r,r'} = \frac{\partial \dot{\lambda}_r}{\lambda_{r'}}
&=-\lambda_r\sum_\mu\sigma_{\mu,r}\sigma_{\mu,r'}h_\mu\frac{2g_\mu^\text{max}}{(1 + C_\mu)^3}\\
 &=-\sum_\mu \sigma_{\mu,r}\left[\frac{2\lambda_r h_\mu g_\mu^\text{max}}{(1 + C_\mu)^3}\right]^\frac{1}{2}\times \sigma_{\mu,r'}\left[\frac{2\lambda_r h_\mu g_\mu^\text{max}}{(1 + C_\mu)^3}\right]^\frac{1}{2}\\
 &= -(A_r)^\top A_r'
\end{split}
\end{align}
where the $r$-th column of $A$ is
\[
A_r^\top = \left[\left(\frac{2\lambda_r h_\mu g_\mu^\text{max}}{(1 + C_\mu)^3}\right)^\frac{1}{2}\sigma_{\mu,r}\right]_{\mu = 1}^{M},
\]
and 
\begin{align}
A= \text{diag}\left( \left[\left(\frac{2\lambda_r h_\mu g_\mu^\text{max}}{(1 + C_\mu)^3}\right)^\frac{1}{2}\right]_{\mu = 1}^{M} \right)\times\Sigma,\quad
J = A^\top A.
\end{align}
We have relied on the assumption that $h_\mu, g_\mu^\text{max} > 0, ~\mu = 1,\dots, M$. Thus $J$ is negative semi-definite, since
\begin{align}
v^\top J v = -v^\top A^\top Av = -(vA)^\top Av \leq 0,
\end{align}
which is a quadratic form. It is strictly negative definite when $A$, and thus $\Sigma$, is full rank. This is only possible when there are utilized fewer non-zero receptors than strains, $R^* \leq M$. Here, the insight from mapping the optimization problem to ecological dynamics is that unless the stringent condition (eq.~\ref{eq_gutimm: general_coexistence}) is met, there will typically be enough unutilized receptors, which contribute rows with only diagonal entries, such that removing them from the matrix will lead to a full rank matrix.

Finally, we show that the objective function (eq. \ref{eq_gutimm: one-type-objective-func}) is a Lyapunov function for the dynamics:
\begin{align}
 \frac{{\rm d}F}{{\rm d}t} &= \sum_{r}\frac{\partial F}{\partial \lambda_r}\frac{{\rm}d\lambda_r}{{\rm d}t} = \sum_{r}-\lambda_r\left[\sum_\mu \sigma_{\mu,r}h_\mu\frac{\partial \tilde{g}_\mu}{\partial C_\mu}(\vec\lambda) + \beta\right]^2 \leq 0.
\end{align}
The time derivative is zero only when $\lambda_r = 0$ or the system reaches a non-zero steady-state.

\subsection{One strain, one receptor}
The immune coverage from the single receptor is given by $C=\sigma \lambda$, and the Lagrangian (eq. \ref{eq_gutimm: general_lagrange}) is,
\begin{align}
L(\lambda, \delta) = h g(\lambda) + \beta \lambda + \delta \cdot(l^2 - \lambda),
\end{align}
with
\[
\tilde{g}(C) = \frac{g^\text{max}}{1 +\sigma\lambda}.
\]
From the KKT conditions, the optimal solution will satisfies,
\begin{align}
0 = \frac{\partial L}{\partial \lambda} &=h\frac{\partial g}{\partial \lambda} +\beta -\delta = -\frac{g^\text{max}\sigma}{(1+\sigma\lambda)^2} + \beta - \delta\\
0 =\frac{\partial L}{\partial \delta} &= l^2 - \lambda.
\end{align}
The second equation imposes the non-negativity constraint as it can only be satisfied when $\lambda \geq 0$. We have two cases to consider. 

\underline{Case 1: $\lambda = 0$.}
Here the inequality constraint is active and $l = 0$ (and $\delta$ may be nonzero), so we check the value of the Lagrangian at this boundary:
\begin{align}
L(\lambda = 0) = h g^\text{max}.
\end{align}
Note that if $h \leq 0$, this is the only viable solution.

\underline{Case 2: $\lambda > 0$.} 
Here the inequality constraint is inactive (i.e. satisfied) and $l^2 > 0$ and $\delta = 0$. From the KKT conditions, we have:
\begin{align}
\label{eq_gutimm: case0_beta_constraint}
\beta + h\frac{\partial g}{\partial \lambda} &= 0,\\
\Rightarrow\beta -\frac{g^\text{max}h\sigma}{(1+\sigma\lambda)^2} &=0.
\end{align}
From this we see that the magnitude of the second term is always less than $g^\text{max}h\sigma$, and since $\beta > 0$, a finite positive solution exists only when $\sigma > \frac{\beta}{g^\text{max}h} > 0$. These equations are solved to find
\begin{align}
\tilde \lambda = \sigma^{-\frac{1}{2}}\left[\left(\frac{g^\text{max}h}{\beta} \right)^{\frac{1}{2}} - \sigma^{-\frac{1}{2}} \right].
\end{align}

The full solution is then,
\begin{align}
\label{eq_gutimm: one-type-optimal-immune-pressure}
\lambda^* = \max\left\{\sigma^{-\frac{1}{2}}\left[\left(\frac{g^\text{max}h}{\beta} \right)^{\frac{1}{2}} - \sigma^{-\frac{1}{2}} \right], 0\right\}
\end{align}
and the explicit $\sigma$ dependence of the solution can be determined by setting $\lambda = 0$ in eq. \ref{eq_gutimm: case0_beta_constraint}:
\begin{align}
\label{eq_gutimm: one-type-leverage}
\text{receptor reactivity threshold:}\quad\sigma^* = \frac{\beta}{g^\text{max}h},
\end{align}
such that $\tilde{\lambda}\vert_{\sigma = \sigma^*} = 0,$ and $\lambda^* = 0$ if $\sigma \leq \sigma^*.$ This means that for a receptor to be effective, there is a threshold level of activity, $\sigma \geq \sigma^*$, that must be met.

\subsection{Two strains, two polyspecific receptors}
We consider the simplest case of a ``cluster'' of immune-interacting strains: two strains and two receptors. The reactivity matrix is assumed to be full-rank\footnote{This excludes certain degenerate solutions that arise when the augmented matrix $[\Sigma^T ~|~ \vec\beta]$ contains redundant rows.} and is given by
\begin{align}
\Sigma &= \begin{bmatrix}
\sigma_{1,1} & \sigma_{1,2}\\
\sigma_{2,1} & \sigma_{2,2}
\end{bmatrix},\\
\begin{split}
\Sigma^{-1} &= \frac{1}{\sigma_{1,1}\sigma_{2,2} - \sigma_{2,1}\sigma_{1,2}}\times\begin{bmatrix}
\sigma_{2,2} & -\sigma_{1,2}\\
-\sigma_{2,1} & \sigma_{1,1}
\end{bmatrix}.
\end{split}
\end{align}
 To find the optimal solution, we must solve, as in eq. \ref{eq_gutimm: one-type-bound}, to equation \ref{eq_gutimm: linear system}:
\begin{align*}
 \Sigma^\top \vec y(\vec\lambda) &= \vec\delta -\beta\boldsymbol{1}.
\end{align*}

\underline{Case 1: $\lambda_1 = \lambda_2 = 0.$}
This is always the optimal solution when $\sum_{\mu \in \{1,2\}} g_\mu^\text{max} h_\mu\sigma_{\mu,r} < \beta$ for $r \in \{1,2\}$. Using eq. \ref{eq_gutimm: general_lagrange}:
\begin{align}
L_\beta(\vec\lambda = 0, \vec\delta) = \sum_{\mu \in \{1,2\}} h_\mu g_\mu^\text{max}
\end{align}
with $l_r = 0, ~r \in \{1,2\}.$

\underline{Cases 2 and 3: $\lambda_1 >0, \lambda_2 = 0$, or $\lambda_1 =0, \lambda_2 > 0.$}
Without loss of generality, consider $\lambda_1 >0, \lambda_2 = 0 $ (i.e $\delta_1 = 0, \delta_2 \neq 0$), after verifying that the reactivity threshold is satisfied (eq. \ref{eq_gutimm: many-type-leverage}), we must solve
\[
\beta = \sum_{\mu \in \{1,2\}} \sigma_{\mu,1}\frac{h_\mu g_\mu^\text{max}} {(1+\sigma_{\mu,1}\lambda_1)^2},
\]
which reduces to a quartic polynomial equation. 

If we assume that $\sigma_{\mu,1} = \sigma >0$ for all strains $\mu$, the equation simplifies to 
\begin{align}
\lambda_1 = \frac{1}{\sigma}\left(\sqrt{\frac{\sigma}{\beta}\times (h_1 g_1^\text{max} + h_2 g_2^\text{max})} - 1\right)
\label{eq_gutimm:coverage-twostrains}
\end{align}
Analogous results hold for the case $\lambda_1 = 0, \lambda_2 > 0 $ (i.e $\delta_1 \neq 0, \delta_2 = 0$).
\subsubsection{Binary binding} 

\underline{Case 4: $\lambda_1, \lambda_2 > 0.$} In this case, when $\Sigma$ is invertible:
\begin{align}
y^* = -(\Sigma^\top)^{-1}\beta\boldsymbol{1} = \frac{-\beta}{\sigma_{1,1}\sigma_{2,2} - \sigma_{2,1}\sigma_{1,2}}\times
\begin{bmatrix}
 \sigma_{2,2} - \sigma_{2,1}\\
 \sigma_{1,1} - \sigma_{1,2}
\end{bmatrix}.
\end{align}
Consider a simpler situation in which, $\sigma_{1,r} = \sigma_{2, r} = \sigma_r \geq 0$ for $r \in \{1,2\}$. That is, strains express the same antigens but the receptors are not necessarily equally reactive. Then, provided $\sigma_1 \neq \sigma_2$, the receptor $j = \arg\max\{\sigma_1, \sigma_2\}$ will out-compete and exclude the other. 

If, without loss of generality, $\sigma_{1,2} = 0$, and $\sigma_{1,1} = \sigma_{2,1},\, \sigma_{2,2} > 0$, we obtain the following simplifications to the previous conditions for co-existence, with
\begin{align}
\vec{y}^* = -(\Sigma^\top)^{-1}\beta\boldsymbol{1} = -\frac{\beta}{\sigma_1\sigma_2}\times
\begin{bmatrix}
 \sigma_{2}-\sigma_1\\
 \sigma_{1}
\end{bmatrix}.
\end{align}
Using eq. \ref{eq_gutimm: one-type-optimal-immune-coverage}, we seek to compute
\begin{align}
\begin{split}
 C_\mu^* &=\left[\frac{g_\mu^\text{max} h_\mu}{|y_\mu^*| }\right]^\frac{1}{2} - 1,
\end{split}
\end{align}
and using eq. \ref{eq_gutimm: strain-steady-noecology},
\begin{equation}
 g_\mu^* = \frac{g_\mu^\text{max}}{1 + C_\mu^*} = \left[\frac{g_\mu^\text{max}|y^*_\mu|}{h_\mu}\right]^\frac{1}{2}.
\end{equation} 
We need to check that the solution is feasible:
\begin{enumerate}
 \item Is harm decreased, i.e. $\vec{y}^* < 0$? For $h_\mu > 0$, this requires that $\sigma_{1} > 0$ and $\sigma_{2} > \sigma_{1}$.

\item Is the immune coverage non-negative, $C_\mu \geq 0$? For strains $\mu \in \{1,2\}$ and receptors $r \in \{1,2\}$, assuming $\sigma_{2} > \sigma_{1}$, we require
\begin{align*}
g_1^\text{max} h_1 \geq |y^*_1|
=\beta\frac{\sigma_2 -\sigma_1}{\sigma_1\sigma_2} = \beta\left(\frac{1}{\sigma_1} - \frac{1}{\sigma_2}\right),\quad g_2^\text{max} h_2 \geq |y^*_2|
=\frac{\beta}{\sigma_2}.
\end{align*}
The second inequality can be rewritten as
\[
\sigma_2 \geq \frac{\beta}{h_2 g_2^\text{max}},
\]
which says that for receptor 2 to be in the repertoire it must be a cost-effective strategy for controlling strain 2, i.e. the potential reduction in harm is greater than the cost. The first inequality can be rearranged to
\[
\sigma_2 \leq \frac{1}{\frac{1}{\sigma_1} - \frac{h_1 g_1^\text{max}}{\beta}},\quad\text{if}\quad \sigma_1 < \frac{\beta}{h_1 g_1^\text{max}},
\]
which says that if receptor 1 is not sufficiently reactive it can be excluded by receptor 2. However, when $\sigma_1 > \frac{\beta}{h_1 g_1^\text{max}}$, the condition holds for any value of $\sigma_2$ and receptor 1 cannot be excluded because it is guaranteed to cost-effectively decrease harm from strain 1. If these conditions are satisfied, then 
\begin{align}
C_1^* =\left[\frac{g_1^\text{max} h_1 \sigma_1\sigma_2}{\beta(\sigma_2 - \sigma_1)}\right]^\frac{1}{2} - 1,\quad C_2^* =\left[\frac{g_2^\text{max} h_2\sigma_2}{\beta }\right]^\frac{1}{2} - 1.
\end{align}

\item Attainable: Next, we use eq. \ref{eq_gutimm: immune-coverage} to find
\begin{align}
 \begin{bmatrix}
  \lambda_1 \\
  \lambda_2
 \end{bmatrix}
 =\Sigma^{-1}\begin{bmatrix}
  C^*_1 \\
  C^*_2
 \end{bmatrix}
 =\frac{1}{\sigma_{1}\sigma_{2}}\times\begin{bmatrix}
   \sigma_{2}C^*_1\\
   \sigma_{1}C^*_2 -\sigma_{1}C^*_2
 \end{bmatrix}.
\end{align}
Finally, we must check that $\lambda_1,\lambda_2 > 0$ (per the KKT conditions). For this to be true, we need to satisfy conditions on the immune coverages:
 \begin{align}
 C_1^* > 0, ~C_2^* > C_1^*.
 \end{align}
which simplifies to two strict inequalities 
\begin{align*}
\begin{split}
&\sigma_2 \leq \frac{1}{\frac{1}{\sigma_1} - \frac{h_1 g_1^\text{max}}{\beta}},\quad\text{if}\quad \sigma_1 < \frac{\beta}{h_1 g_1^\text{max}},\\ &\sigma_2 > \sigma_1\left(1+\frac{h_1 g_1^\text{max}}{h_2 g_2^\text{max}}\right),
\end{split}
\end{align*}
where the second condition is required for receptor 2 to be recruited.
\end{enumerate}

Collectively, there are 3 conditions that must be satisfied for coexistence:
\begin{align}
\begin{split}
&\sigma_2 \geq \frac{\beta}{h_1 g_1^\text{max}},\\
 &\sigma_2 > \sigma_1\left(1+\frac{h_1 g_1^\text{max}}{h_2 g_2^\text{max}}\right),\\
 &\sigma_2 < \frac{\sigma_1}{1 - \frac{h_2 g_2^\text{max}}{\beta}\sigma_1},\quad\text{if}\quad \sigma_1 < \frac{\beta}{h_2 g_2^\text{max}},
\end{split}
\end{align}

\subsubsection{Unconstrained binding}
If $\sigma_{\mu,r}$ are allowed to take any non-negative, possibly bounded value, case 4 becomes more complicated.

\underline{Case 4: $\lambda_1, \lambda_2 > 0.$}
Solving equation \ref{eq_gutimm: linear system} for the case of two and two strains receptors, 
\begin{align}
y^* = -(\Sigma^\top)^{-1}\beta\boldsymbol{1} = \frac{-\beta}{\sigma_{1,1}\sigma_{2,2} - \sigma_{2,1}\sigma_{1,2}}\times
\begin{bmatrix}
 \sigma_{2,2} - \sigma_{2,1}\\
 \sigma_{1,1} - \sigma_{1,2}
\end{bmatrix}.
\end{align}

Recall that $y_\mu(C_\mu)$ is the rate at which harm is decreased by mitigating strain $\mu$, $C_\mu$ is the immune coverage of strain $\mu$, and $|y_\mu| \propto \frac{1}{C_\mu}$. Next we check that the solution is feasible:
\begin{enumerate}
 \item Is harm decreased? For $h_\mu > 0$, this requires that $\sigma_{2,2} \neq \sigma_{2,1}$ and $\sigma_{1,1} \neq \sigma_{1,2}$ and 
 \begin{align*}
  sgn(\sigma_{1,1}\sigma_{2,2} - \sigma_{2,1}\sigma_{1,2}) &= sgn(\sigma_{2,2} - \sigma_{2,1}),\\
  sgn(\sigma_{1,1}\sigma_{2,2} - \sigma_{2,1}\sigma_{1,2}) &= sgn(\sigma_{1,1} - \sigma_{1,2}).
 \end{align*}
 By fixing the sign of the determinant and checking cases, we find that this requires either 
 \[
 \sigma_{2,2} > \sigma_{2,1}, \sigma_{1,1} > \sigma_{1,2}~~\text{or}~~\sigma_{2,2} < \sigma_{2,1}, \sigma_{1,1} < \sigma_{1,2},
 \]
 that is, no single receptor is more reactive than the other to both strains.

\item Is the required rate of decrease achievable? According to equation \ref{eq_gutimm: one-type-bound}, for strains $\{1, 2\}$ and receptors $\{1,2\}$, we require
\begin{align*}
g_1^\text{max}h_1 \geq |y^*_1|
=\beta\frac{\sigma_{2,2} -\sigma_{2,1}}{\sigma_{2,2}\sigma_{1,1} - \sigma_{1,2}\sigma_{2,1}},\quad g_2^\text{max} h_2 \geq |y^*_2|
=\beta\frac{\sigma_{1,1} -\sigma_{1,2}}{\sigma_{2,2}\sigma_{1,1} - \sigma_{1,2}\sigma_{2,1}}
\end{align*}
Assuming $\sigma_{2,2} > \sigma_{2, 1}$ and $\sigma_{1,1} > \sigma_{1, 2}$,
\begin{align*}
\beta\frac{\sigma_{2,2} -\sigma_{2,1}}{\sigma_{2,2}\sigma_{1,1} - \sigma_{1,2}\sigma_{2,1}}&= \beta\frac{\sigma_{2,2} -\sigma_{2,1}}{\sigma_{2,2}\sigma_{1,1} - \sigma_{2,1}\sigma_{1,1} + \sigma_{2,1}\sigma_{1,1}- \sigma_{1,2}\sigma_{2,1}} \\
 &= \frac{\beta}{\sigma_{1,1} + \sigma_{2,1}\frac{\sigma_{1,1}- \sigma_{1,2}}{\sigma_{2,2} -\sigma_{2,1}}},
 \end{align*}
which implies
\begin{align}
 \sigma_{1,1} + \sigma_{2,1}\frac{\sigma_{1,1}- \sigma_{1,2}}{\sigma_{2,2} -\sigma_{2,1}}\ \geq \frac{\beta}{g_1^\text{max} h_1}.
\end{align}
This sets a threshold on the weighted reactivity of the receptor most reactive to strain $1$, accounting for polyspecificity. If $\sigma_{2,1} = 0$, the expression simplifies to
\[
\sigma_{1,1} \geq \frac{\beta}{g_1^\text{max} h_1}.
\]
By symmetry, similar expressions hold for $|y^*_2|$. 

\item Attainable: If the last two conditions are satisfied, we may use eq. \ref{eq_gutimm: one-type-optimal-immune-coverage} to solve for $C_\mu^*$, and find
\begin{align}
 \begin{bmatrix}
  \lambda_1 \\
  \lambda_2
 \end{bmatrix}
 =\Sigma^{-1}\begin{bmatrix}
  C^*_1 \\
  C^*_2
 \end{bmatrix}
 =\frac{1}{\sigma_{1,1}\sigma_{2,2} - \sigma_{2,1}\sigma_{1,2}}\times\begin{bmatrix}
  \sigma_{2,2}C^*_1 - \sigma_{1,2}C^*_2 \\
  \sigma_{1,1}C^*_2 - \sigma_{2,1}C^*_1
 \end{bmatrix}.
\end{align}
Finally, we must check that $\lambda_1,\lambda_2 > 0$ (per the KKT conditions). For this to be true, we need to satisfy conditions on the immune coverages:
\begin{itemize}
 \item If $\sigma_{1,1}\sigma_{2,2} - \sigma_{2,1}\sigma_{1,2} > 0$:
 \begin{align}
 \frac{C_2^*}{C_1^*} <\frac{\sigma_{2,2}}{\sigma_{1,2}}, ~\frac{C_2^*}{C_1^*} >\frac{\sigma_{2,1}}{\sigma_{1,1}},
 \end{align}
 \item If $\sigma_{1,1}\sigma_{2,2} - \sigma_{2,1}\sigma_{1,2} < 0$:
 \begin{align}
 \frac{C_2^*}{C_1^*} >\frac{\sigma_{2,2}}{\sigma_{1,2}}, ~\frac{C_2^*}{C_1^*} <\frac{\sigma_{2,1}}{\sigma_{1,1}}.
\end{align}
\end{itemize}
In two dimensions, this defines a subset of vectors. Going forward, without loss of generality, assume $\sigma_{1,1}\sigma_{2,2} - \sigma_{2,1}\sigma_{1,2} > 0$, then this subset gets bigger as
\[
\frac{\sigma_{1,2}}{\sigma_{2,2}}, \frac{\sigma_{2,1}}{\sigma_{1,1}} \to 0,
\]
that is, as the receptors become less polyspecific.
\end{enumerate}

Towards understanding this condition, suppose $\frac{g_\mu^\text{max} h_\mu}{y_\mu^*} \gg 1$ (i.e. immune coverage $C_\mu^* \gg 1$), then from eq. \ref{eq_gutimm: one-type-optimal-immune-coverage}
\begin{align}
\begin{split}
 C_\mu^* &=\left[\frac{g_\mu^\text{max} h_\mu}{|y_\mu^*| }\right]^\frac{1}{2} - 1
 \approx \left[\frac{g_\mu^\text{max} h_\mu}{|y_\mu^*| }\right]^\frac{1}{2},
\end{split}
\end{align}
and from eq.~\ref{eq_gutimm: strain-steady-noecology},
\begin{equation}
 g_\mu^* = \frac{g_\mu^\text{max}}{1 + C_\mu^*} = \left[\frac{g_\mu^\text{max}|y^*_\mu|}{h_\mu}\right]^\frac{1}{2}.
\end{equation}
Then 
\begin{align}
 \frac{C_2^*}{C_1^*} \sim \left[\frac{g_2^\text{max} h_2 (\sigma_{2,2}-\sigma_{2,1})}{ g_1^\text{max} h_1 (\sigma_{1,1}-\sigma_{1,2})}\right]^{\frac{1}{2}},
\end{align}
and we see that the repertoire is sensitive to the fine details of receptor affinity.

In the most general case, the optimal repertoire is determined by comparing the solutions from all four cases to find the global minimizer. Since convexity is guaranteed when $\Sigma$ is full rank, the optimal solution is case 4 when that solution exists.

\subsection{Many strains and receptors}

\subsubsection{Strains with private receptors}
In general, each receptor may bind to many different strains; however as a simplification, we assume that each strain $\alpha$ is recognized exclusively by a distinct set of receptors $\kappa_\alpha =\{r_{\alpha_1},\dots,r_{\alpha_l}\}\subset \{1,\dots,R\} = \bigcup_{\alpha}\kappa_{\alpha}$. The population of each strain can thus be controlled independently of all others, and the KKT conditions for its associated receptors will be decoupled from all others, such that
\begin{align}
 \begin{split}
  \frac{\partial L}{\partial \lambda_r} &= \sigma_{\alpha,r}h_\alpha\frac{\partial \tilde{g}_\alpha}{\partial C_\alpha}(\vec\lambda) + \beta -\delta_r=0 ~\Rightarrow ~h_\alpha\frac{\partial \tilde{g}_\alpha}{\partial C_\alpha}(\vec\lambda) = \frac{\delta_r -\beta}{\sigma_{\alpha,r}},\\
  \frac{\partial L}{\partial \delta_r} &= l_r^2 -\lambda_r = 0,
 \end{split}
\end{align}
for $r \in \kappa_\alpha$. The RHS of the implication from first equation is identical for all receptors. For the population of a receptor $\lambda_r$ to be greater than zero, we must have $\delta_r = 0$. This cannot be the case for any two receptors, $\lambda_r$ and $\lambda_j$ except if $\sigma_{\alpha,r} = \sigma_{\alpha,j}$. To see this, first observe that if two receptors, $r$ and $j$, coexist they must satisfy the KKT conditions with $\delta_r=\delta_j = 0$, and
\[
\frac{\partial \tilde{g}_\alpha(\vec\lambda)}{\partial C_\alpha}=-\frac{\beta}{\sigma_{\mu,r}} = -\frac{\beta}{\sigma_{\mu,j}}.
\]
Without loss of generality, suppose $\frac{\partial L}{\partial \lambda_j} = 0$ with $\delta_j = 0$, and $\sigma_{\mu,r} > \sigma_{\mu,j}$ then
\[
h_\alpha\frac{\partial \tilde{g}_\alpha(\vec\lambda)}{\partial C_\alpha} = -\frac{\beta_j}{\sigma_{\mu,j}}
\]
and
\[
  0 = \frac{\partial L}{\partial \lambda_{r}} = -\sigma_{\alpha,r}\frac{\beta}{\sigma_{\mu,j}} + \beta -\delta_r.
\]
and since the sum of the first two terms on the RHS is negative, $\delta_r < 0$. This means that we can decrease harm by increasing $\lambda_{r}$ so this cannot be a local minimum, and the repertoire containing $\lambda_j$ is suboptimal. Thus, we only need to select the receptor(s) that satisfy the the leverage cutoff (eq. \ref{eq_gutimm: one-type-leverage}) with the smallest cost-to-affinity ratio
\begin{align}
j_\mu \in \arg\min_{r \in \kappa_\mu}\left\{\frac{\beta}{\sigma_{\mu,r}}\right\},
\end{align}
for each of the $M$ strains. If no receptor meets the leverage cutoff, then we set all their abundances to zero.

\subsubsection{Polyspecific receptors exclude monospecific receptors}
From the the KKT conditions (Eq \ref{eq_gutimm: one-type-KKT}), we can conclude that if there exists a positive solution $\lambda_r^* > 0$ (and $\delta_r = 0)$, i.e., inactive inequality constraint, then
\begin{align}
\begin{split}
\label{eq_gutimm: many-type-leverage}
 0< \beta &= \sum_\mu h_\mu \sigma_{\mu,r}\left[\frac{g_\mu^\text{max}}{(1+C_\mu(\vec\lambda))^2} \right] \\
 &\leq \sum_\mu g_\mu^\text{max} h_\mu\sigma_{\mu,r},
\end{split}
\end{align}
where the LHS is the marginal cost of maintaining receptor $r$ and the RHS of the inequality is the maximum rate of decrease in harm that can be effected by receptor $r$. Thus if this condition is not satisfied for receptor $r$ its optimal value must be $\lambda_r^* = 0.$ This means that a receptor should only be present if the total maximum marginal protection it confers is greater than its cost.

Next, consider a receptor $j$ that binds a single strain $\mu'$, $\sigma_{\mu,j} = \delta_{\mu',\mu}$ and another receptor $r$ that binds the same strain and possibly others, $m_r =\{\mu\} \ni \mu', \sigma_{\mu,r} = \sigma_r\boldsymbol{1}_{\mu \in m_r}$. Suppose receptor $r$ belongs to the optimal repertoire then
\[\beta = \sigma_r\sum_\mu\boldsymbol{1}_{\mu \in m_r}\cdot (-y^*_\mu) = \sigma_r\sum_\mu\boldsymbol{1}_{\mu \in m_r}\cdot |y^*_\mu| \Rightarrow |y_{\mu'}| = \frac{\beta}{\sigma_r} - \sum_{\mu \neq \mu'}\boldsymbol{1}_{\mu \in m_r}\cdot |y^*_\mu| >0\]
where we have used the fact that $y^*_\mu < 0$. For receptor $j$ to also belong to the optimal repertoire, we must have
\[
\sigma_{\mu,j}|y^*_{\mu'}| = \beta \Rightarrow \sigma_{\mu,j} = \frac{\beta}{|y^*_{\mu'}| }= \frac{\beta}{\frac{\beta}{\sigma_r} - \sum_{\mu \neq \mu'}\boldsymbol{1}_{\mu \in m_r}\cdot |y^*_\mu| },
\]
from which we see that increased broad reactivity of receptor $r$ makes it so that receptor $j$ needs to bind more strongly to be in the optimal repertoire. In this way receptors that recognize many strains tend to exclude receptors that are more selective binders.

For a receptor $r$, consider another receptor $j$ that only recognizes strains that receptor $r$ binds to. Suppose $\lambda_r>0$ is part of the optimal repertoire. Then from Eq \ref{eq_gutimm: linear system}, we have
\[
\boldsymbol{\sigma}_r^\top\boldsymbol{y^*} = -\beta
\]
for some $\boldsymbol{y^*} < 0$. If $\boldsymbol{\sigma_{r}} \geq \boldsymbol{\sigma_{j}}$ and $\sigma_{\mu,r} > \sigma_{\mu,j}$ for at least one strain $\mu$,
\begin{align*}
  0 &= \frac{\partial L}{\partial \lambda_{j}} = \boldsymbol{\sigma_{j}}^\top \boldsymbol{y}^* + \beta -\delta_j \\
  &= (\boldsymbol{\sigma_{j}} -\boldsymbol{\sigma_{r}})^\top\boldsymbol{y}^* + \boldsymbol{\sigma_{r}}^\top \boldsymbol{y}^* +\beta- \delta_j\\
  &= (\boldsymbol{\sigma_{j}} -\boldsymbol{\sigma_{r}})^\top\boldsymbol{y}^* -\beta +\beta- \delta_j\\
  &= (\boldsymbol{\sigma_{j}} -\boldsymbol{\sigma_{r}})^\top\boldsymbol{y}^* - \delta_j
\end{align*}
and since the first term is positive, $\delta_j > 0$. This means that we can decrease the objective function by decreasing $\lambda_j$, and so receptor $j$ cannot be in the optimal repertoire if $\lambda_r >0$ is in the repertoire. In this way, polyspecific receptors tend to exclude narrowly reactive ones.

\subsection{Limited ecological interactions}
We consider a system with two strains and one-sided competitive interactions where immune coverage is identical for both strains:
\begin{align}
\tilde{g}_1(C) = \frac{g_1^\text{max}}{1 + C},~~
\tilde{g}_2(C) =\frac{g_2^\text{max}}{1 + C + s\tilde{g}_1(C)}.
\end{align}
For any combination of immune coverage ($C$) and ecological interaction ($s$), using eq.~\ref{eq:delta_strain_clr}, the composition change gap between the two strains induced by immune coverage is given by:
\begin{align}
\begin{split}
\Delta\text{CLR}_1 - \Delta\text{CLR}_2&= \log_{10}(\tilde{g}_1(C)/\tilde{g}_1(0)) - \log_{10}(\tilde{g}_2(C)/\tilde{g}_2(0))\\
&= -\log_{10}\left(\frac{1 + sg^\text{max}_1}{1 + sg^\text{max}_1/(1 + C)^2} \right) \leq 0.
\label{eq:eco_gap}
\end{split}
\end{align}

Qualitatively similar results are obtained with a different model of the effect of competition on the two strains:
\begin{align}
\tilde{g}_1(C) = \frac{g_1^\text{max}}{1 + C},~~
\tilde{g}_2(C) =\frac{g_2^\text{max}}{1 + C}\frac{1}{1 + s\tilde{g}_1(C)},
\end{align}
where
\begin{align}
\Delta\text{CLR}_1 - \Delta\text{CLR}_2 = -\log_{10}\left(\frac{1 + sg^\text{max}_1}{1 + s\tilde{g}_1(C)} \right) \leq 0.
\end{align}


\subsection{Numerical simulations}
To obtain optimal repertoire allocations numerically, for a fixed reactivity matrix ($\Sigma$), eq. \ref{eq_gutimm:cost_function} was solved using the \textit{scikit-learn} package in Python \cite{scikit-learn}. 


\section{Inferring cluster harmfulness}
Generalizing eq.~\ref{eq_gutimm:coverage-twostrains} to an arbitrary number of strains within a cluster co-reactive to the same T~cell, we obtain
\begin{align}
C^*\approx \max\left\{\left[\frac{\sigma\sum_\mu h_\mu g_\mu^\text{max} }{\beta}\right]^{\frac{1}{2}} - 1,~ 0\right\}.
\end{align}
Recall that $K_C$ is the half-maximum constant of inhibition of microbial abundance by repertoire coverage eq.~\ref{eq_gutimm: strain-steady-noecology}, and that our expression for $C^*$ is units of this quantity. Then, assuming the observed cluster coverage, $\langle C_\mu\rangle_k = K_CC^*$ and taking $\sigma \mapsto \sigma/K_C$, we can relate it to the abundance-weighted harm of strains in cluster $k$:
\begin{align}
\sum_{\mu \in \text{cluster } k} h_\mu g_\mu^\text{max}/\sum_{\mu \in \text{cluster } k} g_\mu^\text{max} = \langle h\rangle_k \approx \beta K_C\frac{\left(\langle C_\mu\rangle_k/K_C + 1\right)^2}{\sigma_k\sum_{\mu \in \text{cluster } k} g_\mu^\text{max}},
\end{align}
where $\beta$ is the marginal cost of immunity, $\sigma_k$ is the constant reactivity of strains within the cluster to a common single receptor, and $g_\mu^\text{max}$ is the maximum niche size of strain $\mu$.

From eqs.~\ref{eq:strain_clr} and \ref{eq_gutimm: strain-steady-noecology}, the compositional change ($\Delta\text{CLR}_k$) from pre- to post-immune coverage follows 
\begin{align}
 \Delta\text{CLR}_k = -\log_{10}(1 + \langle C_\mu\rangle_k/K_C) + \frac{1}{9}\sum_{k=0}^8 \log_{10}(1 + \langle C_\mu\rangle_k/K_C),
\end{align}
where the second term is a constant in a regression.

Additionally, our model predicts that sensitivity to engineered changes in repertoire coverage is a proxy for harmfulness: harmful strains will be under tight immune control while commensals may be less controlled (eqs.~\ref{eq_gutimm: strain-steady-noecology}, ~\ref{eq_gutimm: one-type-optimal-immune-pressure}). We note that empirically testing this idea will require additional experiments. Engineering a gradual increase of repertoire coverage is expected to have a more pronounced effect on the abundance of commensals than on harmful strains. Following eq.~\ref{eq_gutimm: one-type-optimal-immune-pressure}, sensitivity to changes in repertoire coverage,
\begin{align}
\label{eq: one-type-optimal-immune-coverage-sensitivity}
\frac{1}{g^\text{max}}\frac{{\rm d}\tilde{g}}{{\rm d}C}\Big|_{C = C^*} &= -\min\left\{\frac{\beta}{\sigma hg^\text{max}},~ \frac{1}{g^\text{max}}\right\},
\end{align}
inversely relates to the protection-to-cost ratio $\frac{\sigma h g^\text{max}}{\beta}$, and can serve as another proxy for microbial harmfulness; the implicit assumption being that more harmful strains already have high immune coverage so their abundance should be less sensitive to changes in coverage (SI Fig.~\ref{SIfig5}B). 

\clearpage{}
\setcounter{figure}{0}
\renewcommand{\thefigure}{S\arabic{figure}}

\begin{figure*}[t!]
\centering
\includegraphics[width=0.9\textwidth]{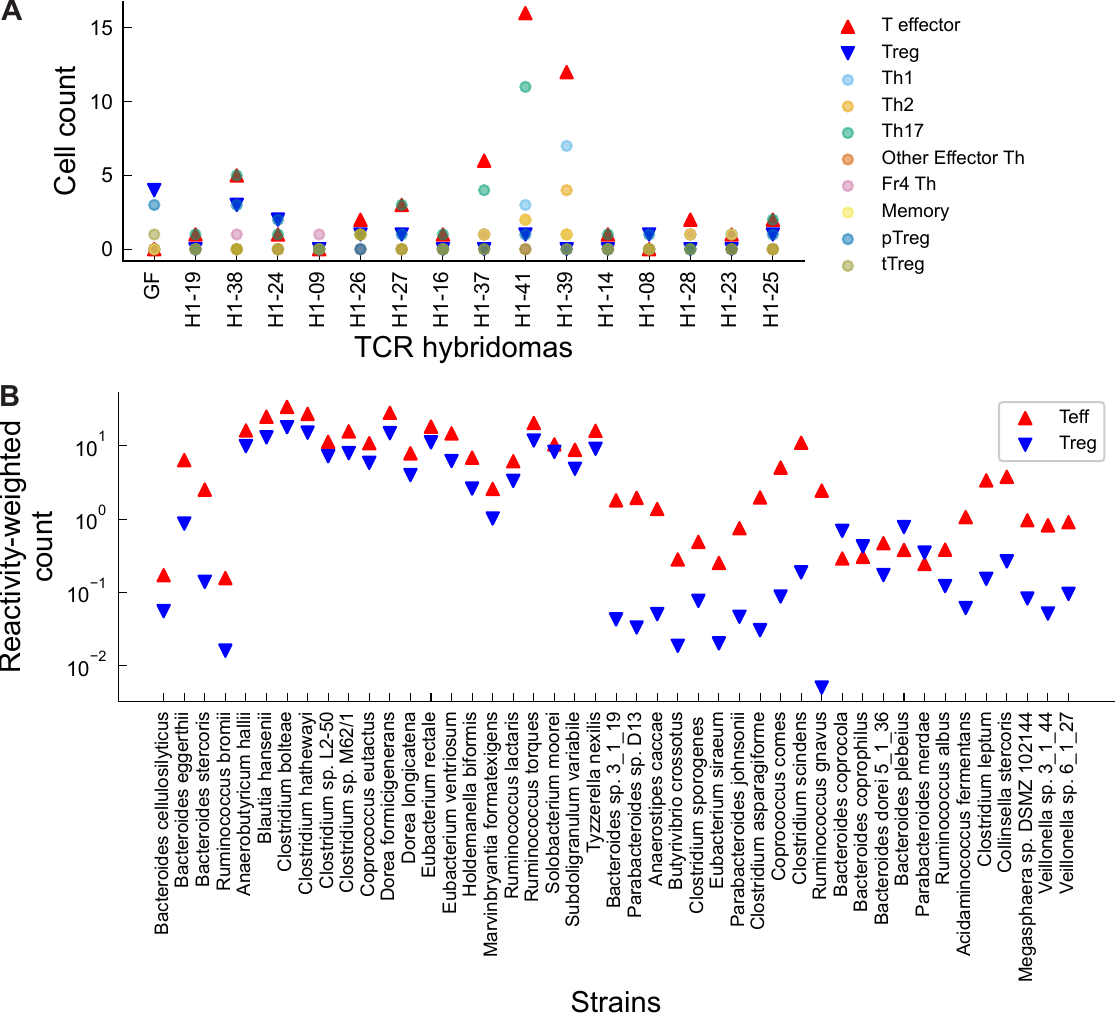}
\caption[T cell repertoire composition]{{\bf T cell repertoire composition.
\label{SIfig1}
} {\bf(A)} Scatter plot showing T~cell subtype counts sampled from the colonic lamina propria of inoculated mice. Colors indicate T~cell subtypes; data from ref.~\cite{Nagashima2023-fe}. {\bf(B)} Effector (red) and regulatory (blue) T~cell reactivity-weighted counts for strains in clusters 1-8.
}
\end{figure*}

\begin{figure*}[t!]
\centering
\includegraphics[width=\textwidth]{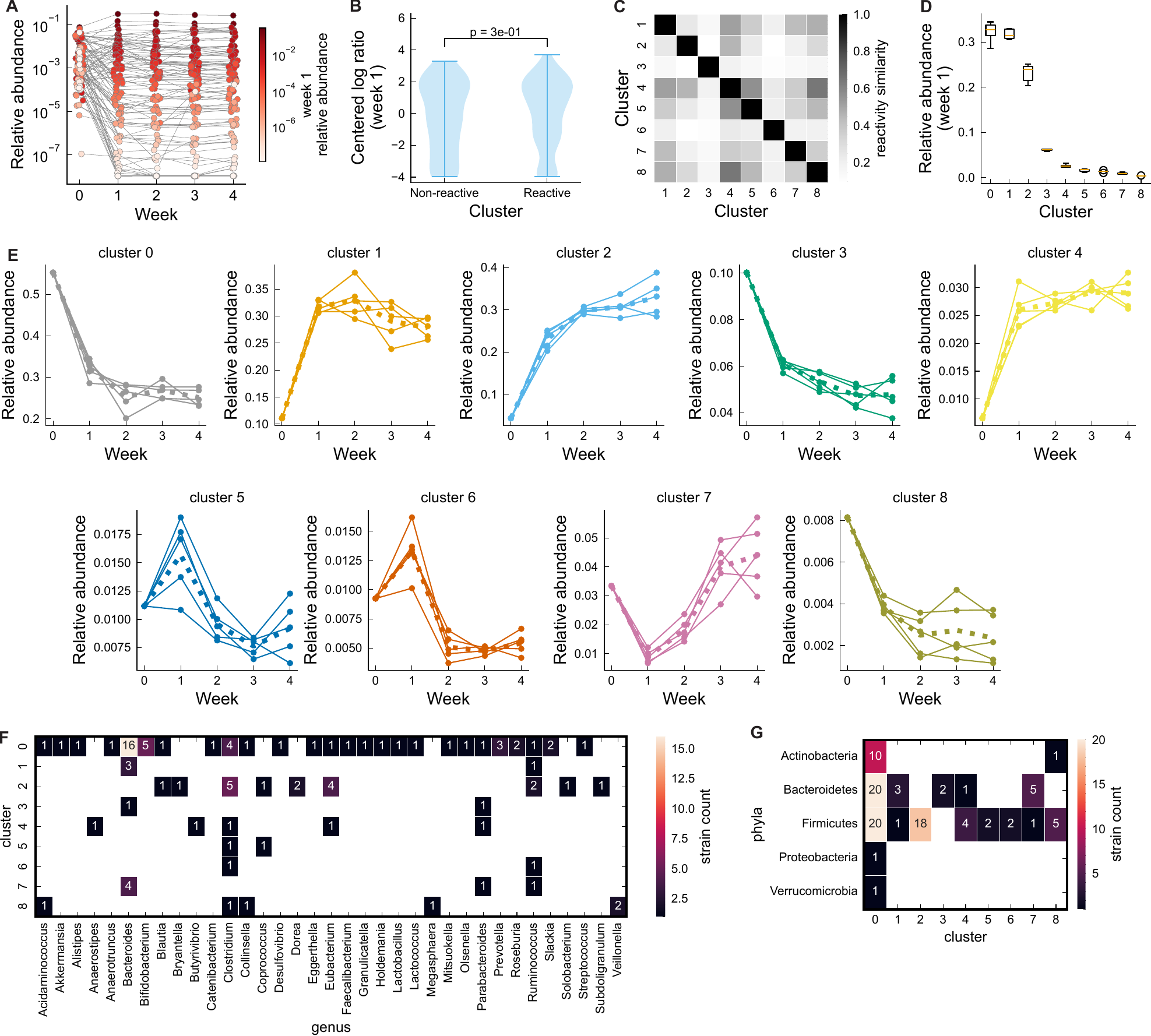}
\caption[Composition of microbiota clusters]{{\bf Composition of microbial clusters.} {\bf(A)} Longitudinal relative abundance, $g_\mu(t)/\sum_\alpha g_\alpha(t)$ of strains $\mu$ in a pre-defined community of 97 microbial strains transplanted into germ free mice; reproduced from ref.~\cite{Wang2023-tc, Nagashima2023-fe}. Trajectories are averaged over 5 biological replicates. Five of the 97 transplanted strains were not detected at week 1 and were dropped from further analysis. Heatmap color indicates strain relative abundance in the microbiota measured at week 1. {\bf(B)} Distribution of week 1 abundances (centered-log ratio) of non-reactive and reactive strains are statistically indistinguishable, as confirmed by a Mann-Whitney test. {\bf(C)} Heatmap showing Bray-Curtis similarity score between clusters, measured by the overlap of normalized T~cell reactivities (Fig.~\mainfigtwo) to each microbial cluster. {\bf(D)} Week 1 relative abundance, $G_k(t=1), ~k=1,\dots,8$, of microbial clusters in the microbiota. Box plots show the mean (orange line) and standard deviation (error bars) across 5 biological replicates. {\bf(E)} Relative abundance of clusters over time (weeks post-inoculation). Solid lines are biological replicates and the dashed line is the mean across replicates. {\bf(F ,G)} Intra-cluster strain abundances across genera (F) and phyla (G).
\label{SIfig2}
}
\end{figure*}

\begin{figure*}[t!]
\centering
\includegraphics[width=\textwidth]{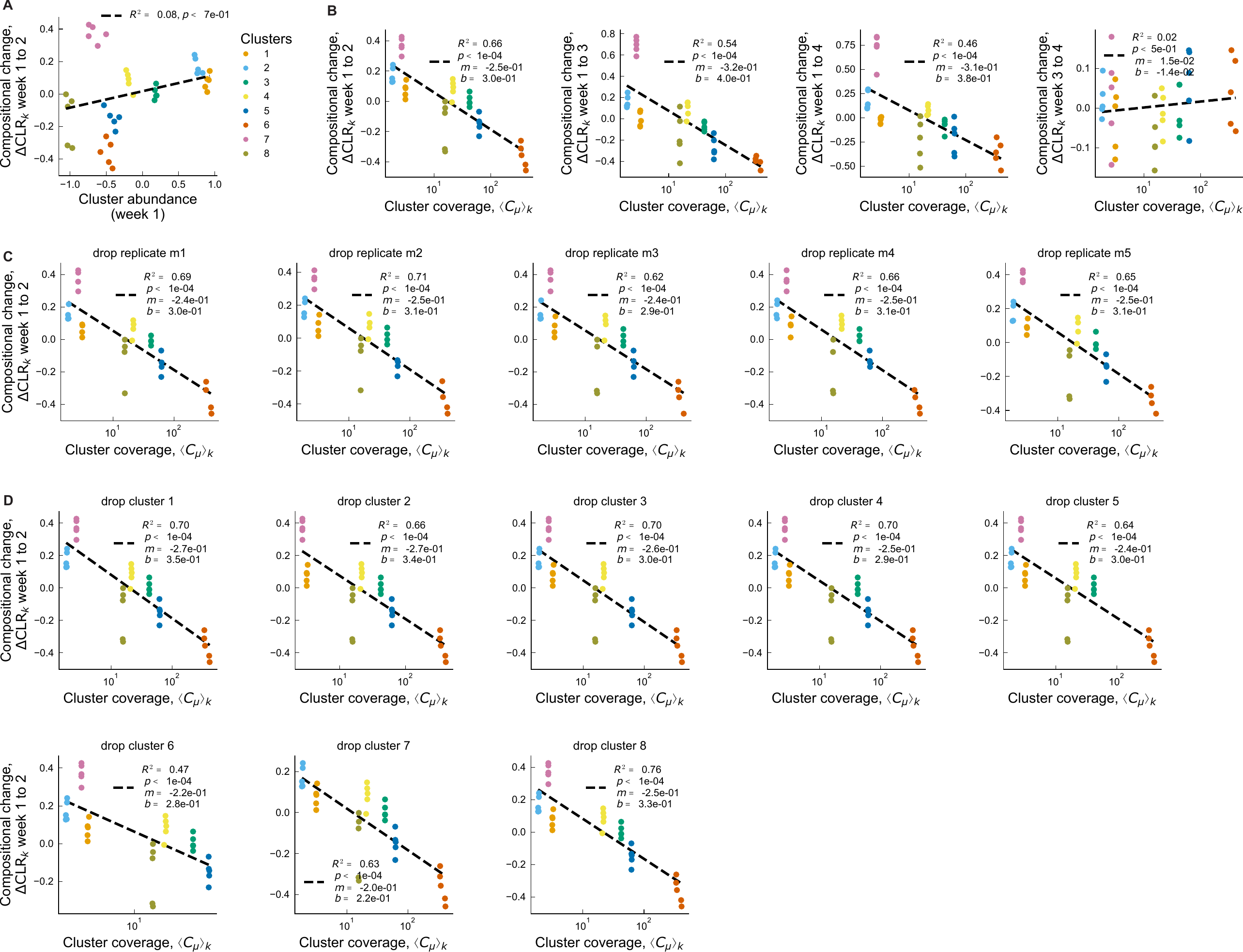}
\caption[Robust relationship between microbiota composition and repertoire coverage]{{\bf Robust relationship between microbiota compositional change and repertoire coverage.
\label{SIfig3}
} {\bf(A)} Scatter plot showing relationship between cluster abundance changes ($\Delta\text{CLR}_k$) from week 1 to 2 versus abundance at week 1 across replicates ($n = 5$; gray dots). Dotted line was fit to: $\Delta\text{CLR}_k = m\cdot\text{CLR}_k(1) + b$. {\bf(B)} Scatter plot showing relationship between abundance changes ($\Delta\text{CLR}_k$) between different time points and repertoire coverage ($\langle C_\mu\rangle_k$) of microbial clusters. Colors represent clusters and points are biological replicates of the same experiment. In each panel, different time points are compared. Dotted line represents fit to $\Delta\text{CLR}_k = m\log_{10}(\langle C_\mu \rangle_k) + b$. Statistical significance was quantified through a hierarchical permutation test (i.e., shuffling cluster and replicate labels). {\bf(C)} Same as (B) but with time fixed and a single indicated replicate dropped to assess robustness of the trend.
{\bf(D)} Same as (B) but with time fixed and a single indicated cluster dropped to assess robustness of the trend. 
}
\end{figure*}

\begin{figure*}[t!]
\centering
\includegraphics[width=0.35\textwidth]{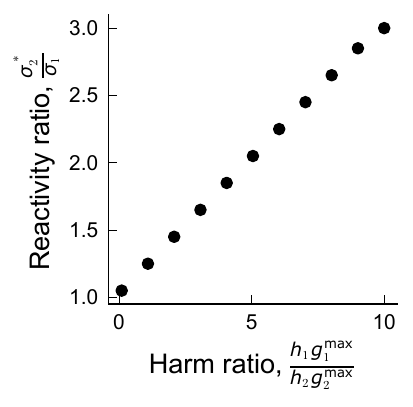}
\caption[Coexistence between a polyspecific and monospecific receptor in the two-strain model]{{\bf Monospecific receptor reactivity threshold for coexistence with a polyspecific receptor in the two-strain model.} Receptor 1 is a polyspecific receptor equally reactive to both strains with reactivities ($\sigma_1 =\sigma_{\mu=1,r=1}=\sigma_{2,1}>0$) while receptor 2 is a monospecific receptor reactive to strain 2 ($\sigma_{2} =\sigma_{2,2} >0;\, \sigma_{1,2}=0$). The plot shows the reactivity ratio of receptor 2 to receptor 1 ($\sigma_2^*/\sigma_1$) required for co-existence, as a function of the harm ratio ($h_1g_1^\text{max}/h_2g_2^\text{max}$) of the two strains. As the potential harm from strain 1 increases, the threshold for including receptor 2 increases ($\sigma_1 = 0.5,~h_2g_2^\text{max} = 5.0$).}
\label{SIfig4}
\end{figure*}

\begin{figure*}[t!]
\centering
\includegraphics[width=0.9\textwidth]{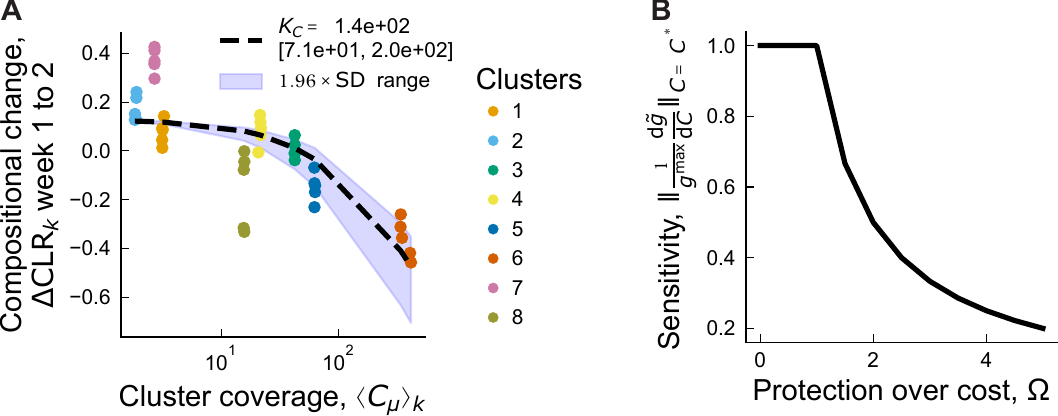}
\caption{{\bf Inferring harmfulness of microbial clusters and strains.
\label{SIfig5}
} {\bf(A)} Scatter plot showing relationship between cluster repertoire coverage ($\langle C_\mu\rangle_k$) and the change in centered log-ratio ($\Delta\text{CLR}_k$) from week 1 to week 2. Points of the same color are biological replicates. Dotted line represents fit to $\Delta\text{CLR}_k = -\log_{10}(1 + K_C^{-1}\langle C_\mu\rangle_k) + d$, and the blue region indicates $\pm 1.96 \times$standard-deviation confidence region for the inferred constant $K_C$. {\bf(B)} Sensitivity of harm-minimized microbe abundance to immune control as a function of the protection-to-cost ratio ($\Omega = \sigma hg^\text{max}\beta^{-1}$). The model predicts that in a well-adapted microbiota, less controlled/commensal species will be more sensitive to changes in their repertoire coverage. Parameters are $\sigma = g^\text{max}= \beta = 1$.
}
\end{figure*}

\begin{figure*}[t!]
\centering
\includegraphics[width=0.4\textwidth]{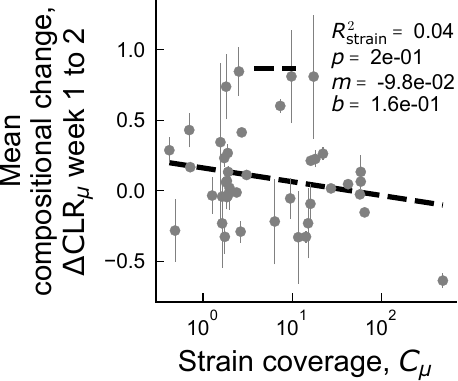}
\caption[Averaging out measurement noise in strain-level data]{{\bf Averaging out measurement noise in strain-level data.
\label{SIfig6}
} Mean of strain composition changes ($\Delta\text{CLR}_\mu$) from week 1 to 2 versus coverage across replicates ($n = 5$; gray dots). Data was fit to: $\text{SD} = m\cdot\text{CLR} + b$. Averaging out measurement noise does not recover cluster-level correlation ($R^2_\text{cluster} = 0.66$; Fig.~\mainfigthree~C) between composition changes and coverage.
}
\end{figure*}

\begin{figure*}[t!]
\centering
\includegraphics[width=0.6\textwidth]{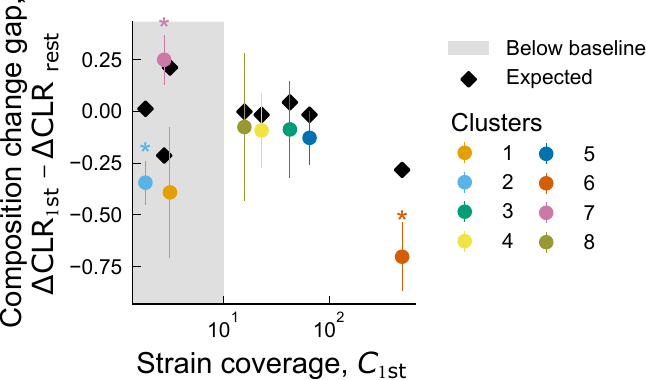}
\caption{{\bf Composition change gap ($\Delta\text{CLR}_{\text{1st}^{(k)}}-\Delta\text{CLR}_{\text{rest}^{(k)}}$) plotted against the immune coverage of the most abundant strain in each cluster ($C_\text{1st}$).
\label{SIfig7}
} The gray shaded region denotes clusters with coverage below the biological baseline ($C^\text{low} = 10$). Colored circles indicate the observed gaps ($\pm 1$ standard deviation). Statistical significance of gaps ($p < 0.05$; *) was determined by a paired t-test of 5 biological replicates with Benjamini-Hochberg correction. If immune coverage alone determined compositional change, then variation in the coverage of individual strains ($C_\mu$; Fig.~\mainfigtwo) would drive predictable variation in compositional change ($\Delta\text{CLR}_\mu$). Black diamonds represent the within-cluster gaps in compositional changes that would be expected if strain compositional changes followed the cluster-level fit in Fig.~\mainfigthree~C: $\Delta\text{CLR} = m\log_{10}(C_\mu) + b$ with $m=-0.25$, and $b=0.3$. The quantity $\Delta\text{CLR}_{\text{rest}^{(k)}}$ is calculated using the abundance-weighted mean immune coverage of all strains in the cluster excluding the most abundant strain.
}
\end{figure*}

\begin{figure*}[t!]
\centering
\includegraphics[width=0.9\textwidth]{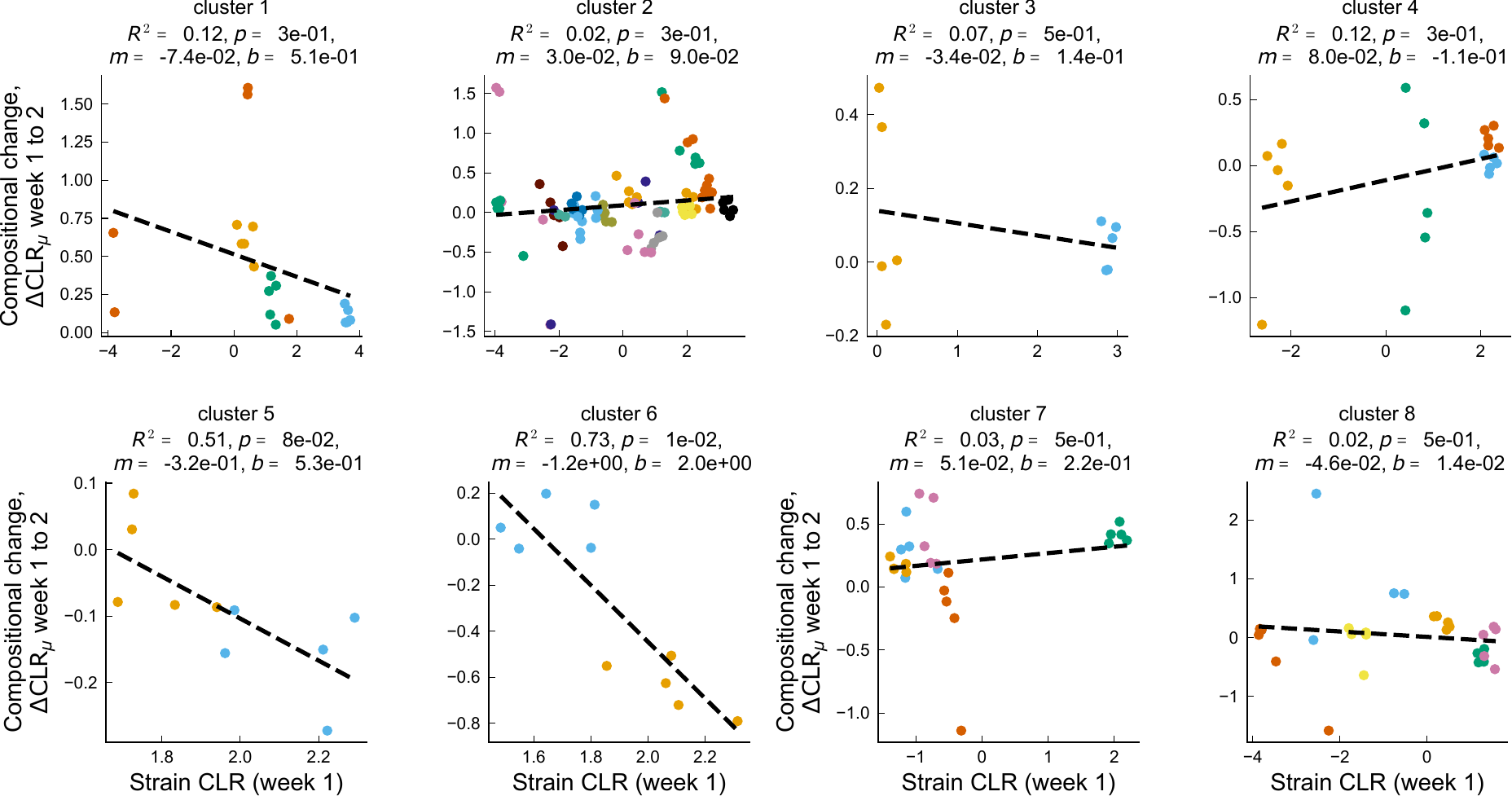}
\caption[Intra-cluster trends in strain compositional changes]{{\bf Intra-cluster trends in strain compositional changes.
\label{SIfig8}
} Relationship between compositional changes ($\Delta\text{CLR}$) and week 1 composition of strains within clusters. Dotted line represents a fit to $\Delta\text{CLR}_\mu(1,2) = -m\cdot\text{CLR}_\mu(1) + b$. Statistical significance ($p$-values) were corrected for multiple testing (Benjamini-Hochberg). 
}
\end{figure*}

\bibliographystyle{plos2015}
\bibliography{supp_references}